\documentclass{emulateapj}  
\usepackage[toc,page]{appendix}

\usepackage{amsmath}
\usepackage{amssymb}
\usepackage{verbatim}
\usepackage{graphicx}
\usepackage{morefloats}
\usepackage{float}
\usepackage{lipsum}
\usepackage{subfigure}
\usepackage{longtable}
\usepackage{lipsum} 
\usepackage{rotating}
\usepackage{wasysym}
\usepackage{natbib}
\slugcomment{Accepted by ApJ}

\newcommand{\sm}{M$_{\odot}$}

\newcommand{\msol}{M$_{\odot}$}

\newcommand{\ergs}{$~{\rm erg~s^{\scriptscriptstyle -1}}$}

\newcommand{\kms}{\ensuremath{{\rm km~s}^{-1}}}

\newcommand{\snaj}{SN~2006aj/GRB~060218}
\newcommand{\snbw}{SN~1998bw/GRB~980425}
\newcommand{\sndh}{SN~2003dh/GRB~030329}
\newcommand{\snlw}{SN~2003lw/GRB031203}

\newcommand{\snbh}{SN 2010bh/GRB~100316D}
\newcommand{\snma}{SN 2010ma/GRB~101219B}

\newcommand{\HeFive}{\ion{He}{1}~$\lambda$5876}
\newcommand{\HeSix}{\ion{He}{1}~$\lambda$6678}
\newcommand{\HeSeven}{\ion{He}{1}~$\lambda$7065}
\newcommand{\FeFive}{\ion{Fe}{2}~$\lambda$5169}
\newcommand{\FeFivezero}{\ion{Fe}{2}~$\lambda$5018}
\newcommand{\FeFour}{\ion{Fe}{2}~$\lambda$4924}

\newcommand{\SiSix}{\ion{Si}{2}~$\lambda$6355}

\newcommand{\OxySeven}{\ion{O}{1}~$\lambda$7774}

\newcommand{\nSNGRB}{11} 
\newcommand{\nSNIcbl}{10}  
\newcommand{\nSNIcbltot}{23}  
\newcommand{\nSNIc}{17}

\newcommand{\nspecSNGRB}{119} 
\newcommand{\nspecSNIcbl}{120} 
\newcommand{\nspecSNIc}{209} %
\newcommand{\nspectotaltotal}{448} 
\newcommand{\nspeclatetime}{41} 
\newcommand{\nspecanal}{407} 

\defcitealias{modjaz14}{M14}

\begin{document}

\title{The Spectral SN-GRB Connection: Systematic Spectral Comparisons between Type Ic Supernovae, and Broad-Lined Type Ic Supernovae with and without Gamma-Ray Bursts}

\author{Maryam~Modjaz\altaffilmark{1}, %
Yuqian~Q.~Liu\altaffilmark{1},
Federica~B.~Bianco\altaffilmark{1},
Or~Graur\altaffilmark{1,2}
}

\altaffiltext{1}{Center for Cosmology and Particle Physics, New York University, 4 Washington Place, New York, NY 10003, USA; mmodjaz@nyu.edu.}
\altaffiltext{2}{Department of Astrophysics, American Museum of Natural History, Central Park West and 79th Street, New York, NY 10024, USA}

\begin{abstract}
We present the first systematic investigation of spectral properties of \nSNIc\ Type Ic Supernovae (SNe Ic), \nSNIcbl\ broad-lined SNe Ic (SNe Ic-bl) without observed Gamma-Ray Bursts (GRBs) and \nSNGRB\ SNe Ic-bl with GRBs (SN-GRBs) as a function of time in order to probe their explosion conditions and progenitors. Using a number of novel methods, we analyze a total of \nspecanal\ spectra, which were drawn from published spectra of individual SNe as well as from the densely time-sampled spectra of \citet{modjaz14}. In order to quantify the diversity of the SN spectra as a function of SN subtype, we construct average spectra of SNe Ic, SNe Ic-bl without GRBs and SNe Ic-bl with GRBs. We find that SN~1994I is not a typical SN Ic, in contrast to common belief, while the spectra of \snbw\ are representative of mean spectra of SNe Ic-bl with GRBs. We measure the ejecta absorption and width velocities using a new method described here and find that SNe Ic-bl with GRBs, on average, have quantifiably higher absorption velocities, as well as broader line widths than SNe without observed GRBs. In addition, we search for correlations between SN-GRB spectral properties and the energies of their accompanying GRBs. Finally, we show that the absence of clear He lines in optical spectra of SNe Ic-bl, and in particular of SN-GRBs, is not due to them being too smeared out due to the high velocities present in the ejecta. This implies that the progenitor stars of SN-GRBs are probably He-free, in addition to being H-free, which puts strong constraints on the stellar evolutionary paths needed to produce such SN-GRB progenitors at the observed low metallicities.
\end{abstract}

\section{Introduction}\label{intro_sec}

Stripped supernovae (SNe) and long-duration Gamma-Ray Bursts (long GRBs) are amongst nature's most powerful explosions from massive stars. They energize and enrich their host galaxies, and, like beacons, they are visible over large cosmological distances. However, their progenitor systems remain to be identified and the details of the explosion mechanism are unknown (see reviews by \citealt{woosley06_rev} and \citealt{smartt09_rev}). Stripped-envelope SNe (i.e, SNe of Types IIb, Ib, and Ic , e.g., \citealt{filippenko97_review,modjaz14}), in short ``stripped SNe," are core-collapse events whose massive progenitors have been removed of progressively larger amounts of their outermost H and He envelopes.

Specifically, broad-lined SNe~Ic (SNe~Ic-bl) are SNe Ic whose line widths approach 15,000$-$30,000 \kms\ around maximum light and whose optical spectra show no obvious presence of H and He. Nevertheless, the exact distinction between a normal SN Ic and a SN Ic-bl is still debated, as well as whether there is hidden He in SN Ic-bl spectra. 

For the last fifteen years, the exciting connection between long GRBs and
stripped SNe has been observationally well established, in particular with SNe Ic-bl, as the only type of SNe observed accompanying long GRBs for both low-luminosity, nearby, and high-luminosity, cosmological GRBs (for reviews, see \citealt{woosley06_rev,hjorth11,modjaz11_rev} and for the most recent SN accompanying a high-luminosity GRB, \citealt{xu13}). However, the existence of many more SNe~Ic-bl \textbf{without} GRBs (most convincingly SN~Ic-bl 2002ap without a relativistic engine, e.g., \citealt{berger02}) raises the question of what distinguishes SN-GRB progenitors from those of ordinary SN~Ic-bl without GRBs. Viewing angle effects are probably not the only reason why those SNe Ic-bl did not show an accompanying GRB \citep{soderberg06_radioobs,soderberg10_09bb}. One promising line of
research is to investigate what sets apart SNe \textbf{with} GRBs from
those \textbf{without} GRBs in order to elucidate the conditions and progenitors of these two types of explosions. While there are numerous possible avenues (for recent reviews, see e.g., \citealt{hjorth11,modjaz11_rev}), here we concentrate on comparing the optical spectra of SNe Ic-bl with GRBs to those without GRBs, as well as to those of normal SNe Ic. Optical, early-time, i.e., photospheric phase, spectra are used for the classification of different explosions and probe the bulk of the ejected stellar material, in particular the outermost layers. Prior work involving synthetic models based on Monte Carlo radiative transfer codes (\citealt{mazzali13,walker14} and references therein), while important, has included only two normal SNe Ic (SN~1994I, SN~2004aw) and only four SN Ic-bl without GRBs, thus not yet providing a statistically significant sample. In particular, in other comparisons in the literature (e.g., \citealt{pian06,valenti08,taubenberger06}), SN~1994I is assumed to represent the class of SNe Ic, while there is strong evidence that SN~1994I is not a ``typical" SN Ic, neither photometrically \citep{richardson06,drout11}, nor spectroscopically, as we will show below.

The time is ripe to undertake this detailed comparison, given that from the literature and as part of the CfA Stripped SN data release (\citealt{modjaz14}, hereafter \citetalias{modjaz14}), we can now include a statistically significant set of \nSNIcbltot\ SNe Ic-bl (with and without GRBs) and \nSNIc\ SNe Ic that have accompanying photometry, so that we can assign phases to the spectra. Now, for the first time, we have a large enough sample to conduct  a systematic, data-driven study on quantifying the similarity and behavior of SNe Ic-bl (especially those accompanied with GRBs) compared to those with SNe Ic. Here, we concentrate on the SN-GRB connection and use the same novel methodology we develop in \citet{liu15}, in which we concentrate on SNe IIb, Ib and Ic.

The format is as follows: in Section~\ref{sample_sec}, we summarize the sample of SNe used in this paper, which we analyze in the remainder of the paper using two complimentary methods.  In Section~\ref{averagespec_sec}, we quantify the spectral diversity of stripped-envelope SNe by analyzing and comparing mean spectra, as well as corresponding standard deviation and maximum deviation spectra, which we construct for SNe Ic, SNe Ic-bl, and SNe Ic-bl connected with GRBs. This method takes into account all features that compose the spectrum. We also use mean spectra to explore the question of whether SNe Ic-bl may have smeared out He lines hidden in their spectra - one of the outstanding questions in understanding the progenitors of SN-GRBs as it is hard to produce a star that is both H- and He-free at the low metallicities at which SN-GRBs are observed.
In Section~\ref{specline_sec}, we concentrate on the velocity evolution of the \FeFive\ line and conduct a comparison of its behavior (both in terms of its Doppler velocity and its width) amongst the SN types, for which we developed a new method for measuring blended lines in spectra of SNe Ic-bl, described in the Appendix. In Section~\ref{discussion_sec}, we discuss our results and their implications for progenitor models of SNe Ic-bl, specifically those with and without observed GRBs, and conclude with Section~\ref{conclusion_sec}.

\section{SN spectral sample and other relevant properties}\label{sample_sec}

\begin{deluxetable*}{llcl}
\tablecolumns{2}
\singlespace
\tablecaption{Sample of SNe Ic }
\tablehead{
\colhead{SN name} &
\colhead{redshift $z$} &
\colhead{Phases of spectra with respect to maximum light\tablenotemark{a}}  &
\colhead{References\tablenotemark{b}}
}
\startdata
SN 1983V\tablenotemark{$\dagger$} & 0.0054 &   -15, -13, -10, -2, 6, 7, 8, 9, 10, 11, 12, 32 & SNID(C97) \\ [1ex]
SN 1990B\tablenotemark{c} &  0.0075  &  4, 5, 6, 6, 7, 9, 10, 15, 22, 28, 28, 31, 39, 47, 58, 65, 72, 88, 90+(2) & SNID(B90, M01, CfA) \\ [1ex]
SN 1992ar\tablenotemark{c} &   0.1450 & 3  & C00 \\ [1ex]
SN 1994I & 0.0015&   -6, -6, -5, -4, -3, -3, -2, -1, 1, 2, 2, 3, 3, 6, 7, 11, 21, 22, 23, 24, & SNID(F95, C96, M99), M14 \\ [1ex]
              &           &  25, 26, 28, 30, 32, 33, 34, 34, 36, 37, 38, 40, 55, 56, 68,90+(3)    & \\  [1ex]
SN 2004aw &  0.0158 &  -5, -3, -2, -2, -1, 0, 1, 2, 2, 3, 3, 4, 5, 12, 18, 18, 19, 21, 23,  & SNID(T06), M14 \\ [1ex]
		   &    	    & 25, 26, 30, 32, 46, 48, 90+(2)   &  \\      [1ex]
SN 2004dn &  0.0126 & -8, 31, 34, 52, 54 & H08, M14\\ [1ex]
SN 2004fe\tablenotemark{$\dagger\dagger$} &  0.0179 &  -7, -7, -6, -5, -1, 0, 1, 2, 3, 9, 22,  90+(1) & M14\\ [1ex]
SN 2004ge & 0.0155 &     12  & M14\\ [1ex] 
SN 2004gt &  0.0055 &   16, 18, 19, 22,  43, 48, 70, 81, 90+(5) & M14\\ [1ex]
SN 2005az &  0.0088 &  -8, -7, -5, -3, -1, 1, 4, 17, 18, 21, 23, 25, 27, 29, 30, 31, 32, & M14\\ [1ex]
                 &            &  48, 56, 61, 64, 86  & \\ [1ex]
SN 2005kl &     0.0035  & -4, 39, 70, 90+(1) & M14\\ [1ex]
SN 2005mf & 0.0189  & -2, -1, 5, 8  & M14\\ [1ex]
SN 2007cl &  0.0222 &    -6, -6 & S12, M14\\ [1ex]
SN 2007gr &   0.0017 &  -9, -8, -6, -5, -4, -3, 0, 5, 7, 8, 9, 10, 13, 14, 15, 16, 17, 18, & V08, M14\\ [1ex]
		   & 	  	   &    20, 25, 42,  45, 47, 49, 53, 53   & \\ [1ex]
SN 2011bm\tablenotemark{$\dagger\dagger\dagger$} &    0.0221  & -11, -3, -1, 16, 24, 30, 50, 61, 90+(5)   & V12\\ [1ex]
PTF12gzk\tablenotemark{$\dagger\dagger\dagger\dagger$}\tablenotemark{d} & 0.0138 & -10, -5, -4, 0, 2, 3, 4, 6 & B12 \\ [1ex]
SN 2013dk &  0.0055 &   -1  & E13
\enddata
\tablenotetext{a}{Phases are in the rest-frame with respect to $V$-band maximum light, either directly measured (as listed in \citet{modjaz14} or in the original papers or using the polynomial fit as described in \citet{bianco14}), or transformed (see text for details). Phases are rounded to the nearest whole day. The number in a bracket is the number of spectra with phases larger than 90 days after the date of maximum light - spectra which we include for completeness, but do not analyze here. 
}
\tablenotetext{b}{References: SNID = in SNID release version 5.0 via templates-2.0 by \citet{blondin07}, with the original references in parentheses;  C97 = \citet{clocchiatti97}; B90 = \citet{barbon90}; M01 = \citet{matheson01}; CfA = spectra from the CfA Supernova Program, published here;  M14 = \citet{modjaz14}; M99 = \citet{millard99}; F95 = \citet{filippenko95}; C96 = \citet{clocchiatti96}; H08 = \citet{harutyunyan08}; T06 = \citet{taubenberger06}; S12 = \citet{silverman12_data}; V12 = \citet{valenti12}; B12 = \citet{benami12}; E13 = \citet{eliasrosa13}. }
\tablenotetext{c}{Note that the dates of maximum for SNe~1990B and 1992ar are uncertain since their observed light curves start after maximum, and their dates were estimated via fitting with the light curve of other well-observed Stripped SNe.}
\tablenotetext{d}{While SNID finds good matches of PTF 12gzk with SN~Ic 2004aw, a SN Ic, we note that it  displays very high absorption velocities ($v_{abs}$, similar to those of SN Ic-bl, as inferred from the optical spectra (e.g., \citealt{benami12}) and from radio observations \citep{horesh13}. Thus, while we call it a "SN Ic" in the table, we are not including PTF12gzk when constructing mean spectra of SN Ic (see Section~\ref{averagespec_sec} for details).}
\tablenotetext{$\dagger$}{For SN~1983V, we determine the date of $V$-band maximum by applying the polynomial fit as described in \citet{bianco14} to the original data of C97, to be consistent with the date of maximum determination for the rest of the SNe in our sample. Our date of maximum is $JD_V$ = 2,445,681.4 $\pm$ 1.0, which is 6 days later than the date obtained in C97 with light curve template fitting. }
\tablenotetext{$\dagger\dagger$}{For SN~2004fe, we use the date of $V$-band maximum as measured by \citet{bianco14}, which is different from that in \citet{drout11}.}
\tablenotetext{$\dagger\dagger\dagger$}{For 2011bm, we determine the date of $V$-band maximum by applying the polynomial fit as described in \citet{bianco14} to the original data of V12, and obtain $JD_V = 2,455,673.7 \pm$ 2.1.   }
\tablenotetext{$\dagger\dagger\dagger\dagger$}{For PTF12gzk, we determine the date of $V$-band maximum by applying the polynomial fit as described in \citet{bianco14} to the original data of B12, and obtain $JD_V = 2,456,150.1 \pm 0.7$. }
\label{table_snIcsample}
\end{deluxetable*}

\begin{deluxetable*}{llcl}
\tablecolumns{2}
\singlespace
\tablecaption{Sample of SNe Ic-bl }
\tablehead{
\colhead{SN name} &
\colhead{Redshift $z$} &
\colhead{Phases of spectra with respect to maximum light\tablenotemark{a}}  &
\colhead{References\tablenotemark{b}}
}
\startdata
SN 1997ef & 0.0113 &    -14, -12, -11, -10, -9, -6, -5, -5, -4, 7, 13, 14, 16, 17, 19,   & I00, M14\\ [1ex]
		   &          & 20, 22, 27, 38, 40, 44, 46, 48, 48, 75, 80, 86, 90+(1)   & \\  [1ex]
SN 2002ap &  0.0022&   6, -5, -2, -1, 0, 1, 2, 3, 4, 5, 5, 6, 7, 7, 10, 12, 13, & SNID(G02, F03), M14 \\ [1ex]
			&       &  26, 27, 30, 31, 31, 90+(12) 	& \\  [1ex]
SN 2003jd &  0.0188 &  -3, -2, -1, 0, 1, 19, 20, 21, 22, 23, 24, 27, 28, 34,  & V08, M14\\ [1ex]
		   & 	          & 46, 47, 48, 49, 49, 51, 54, 55, 60, 73, 90+(1)  & \\   [1ex]
SN 2007D &   0.0232 &   -8  & M14\\ [1ex]
SN 2007bg\tablenotemark{$\dagger$} &  0.0346 &  6, 7, 19, 27, 59  & Y10, M14 \\ [1ex]
SN 2007ru & 0.0155 &  -3, 0, 4, 5, 9, 13, 16, 29, 36, 59, 90+(1) & S09, M14\\ [1ex]
SN 2010ay\tablenotemark{$\dagger$} & 0.0671 & 16, 26 & Sa12 \\ [1ex]
PTF10bzf (SN~2010ah)\tablenotemark{$\dagger$} &  0.0498 &  2, 7 & C11\\ [1ex]
PTF10qts\tablenotemark{$\dagger$}  & 	0.0907	&   -1, 1, 10, 17, 20, 24, 90+(1)  	& W14 \\ [1ex]
PTF10vgv\tablenotemark{$\dagger$}\tablenotemark{c}   & 0.0142 &  -7, 4, 8, 37, 73   & C12  \\ [1ex]
SN 1998bw/GRB~980425\tablenotemark{$^{\ast}$} & 0.0085 &  -11, -9, -8, -5, -4, -3, -1, 1, 2, 4, 7, 9, 10, 11, 17,  & G98, P01 \\ [1ex]
					&    &  20, 27, 43, 50, 62, 71, 90+(5) & \\   [1ex]
SN 2003dh/GRB~030329 & 0.1685 &   -10, -9, -8, -7, -6, 6, 12, 14, 18, 30, 56 & S03, M03, K03, K04, D05 \\ [1ex]
SN 2003lw/GRB~031203\tablenotemark{$\dagger$}\tablenotemark{$^{\ast}$} & 0.1006 &    -2, 7 & M04\\ [1ex]
SN 2006aj/GRB~060218\tablenotemark{$^{\ast}$} &  0.0335 & -7, -6, -6, -5, -5, -4, -3, -3, -2, -2, -1, -1, 0, 0, & M06, P06, M14 \\ [1ex]
					&  	& 1, 2, 2, 3, 3, 4, 5, 6, 8, 9, 10, 90+(1)  & \\  [1ex]
SN~2009bb\tablenotemark{d} & 0.0099 &   -5, -4, 1, 5, 15, 16, 20, 21, 24, 28, 42 & P11\\ [1ex]
SN~2009nz/GRB~091127\tablenotemark{e}   & 0.490 & $\sim$ 1 & B11 \\ [1ex]
SN 2010bh/GRB~100316D\tablenotemark{$\dagger$}\tablenotemark{$^{\ast}$} &   0.0593 & -4, -4, -4, -4, -3, -2, -1, 0, 0, 1, 1, 2, 4, 5, 6, & C10, B12 \\ [1ex]
												&       & 8, 10, 12, 14, 15, 17, 24, 30, 33   & \\   [1ex]
SN 2010ma/GRB~101219B &  0.55 &   -13, -3, 10 & S11\\ [1ex]
SN 2012ap\tablenotemark{d}  & 0.0121   & -5, -3, -1, 1, 2, 4, 7, 8, 14, 25  & M15 \\ [1ex]
SN 2012bz/GRB~120422A &  	0.2825 &  1, 4, 9 & S14\\ [1ex]
SN 2013cq/GRB~130427A\tablenotemark{$\dagger$} &   	0.3399 &  -1 & X13 \\   [1ex]
SN~2013dx/GRB~130702A & 0.1450 & -5, -5, -3, -2, -1, 1, 1, 2, 5, 5, 6, 11, 13, 17, 20, 23    & D15 \\ [1ex]
SN 2013ez/GRB~130215A\tablenotemark{f} & $>0.597$ & $\sim$0 & C14 \\ [1ex]
\enddata
\tablenotetext{a}{Phases are in the rest-frame with respect to $V$-band maximum light, either directly measured (as listed in \citet{modjaz14} or in the original papers or using a polynomial fit as described in \citet{bianco14}), or transformed (see text for details). Phases are rounded to the nearest whole day. The number in a bracket is the number of spectra with phases larger than 90 days after the date of maximum light - spectra which we include for completeness, but do not analyze here.
}
\tablenotetext{b}{References:  I00 = \citet{iwamoto00}; M14 = \citet{modjaz14}; SNID = in SNID release version 5.0 via templates-2.0 by \citet{blondin07}, with the original references in parentheses;; G02 = \citet{gal-yam02_02ap}; F03 = \citet{foley03}; V08 = \citet{valenti08}; Y10 = \citet{young10}; S09 = \citet{sahu09}; Sa12 = \citet{sanders12}; C11 = \citet{corsi11}; C12 = \citet{corsi12}; W14 = \citet{walker14}. G98 = \citet{galama98}; P01 = \citet{patat01}; S03 = \citet{stanek03}; M03 = \citet{matheson03}; K03 = \citet{kawabata03}; K04 = \citet{kosugi04} D05 = \citet{deng05}; M04 = \citet{malesani04}; M06 = \citet{modjaz06}; P06 = \citet{pian06}; P11 = \citet{pignata11}; C10 = \citet{chornock10}; B12 = \citet{bufano12}; S11 = \citet{sparre11}; M15 = \citet{milisavljevic15}; S14 = \citet{schulze14}; X13 = \citet{xu13}; C14=\citet{cano14}. D15 = \citet{Delia15_sngrb13dx}.}
\tablenotetext{c}{While \citet{corsi12} classify PTF10vgv a SN Ic based on its measured low \SiSix\ absorption velocities, its optical spectra show broad lines amd are clearly those of a SN Ic-bl, with SNID finding good matches with only SNe Ic-bl. Thus we re-classify this object as a SN Ic-bl, following our classification approach outlined in M14. }
\tablenotetext{d}{We include SNe~2009bb and 2012ap as "engine-driven" SNe Ic-bl, since they have been suggested to be relativistic SNe based on their large radio-emission, similar to, albeit weaker than, nearby SN-GRBs, even if no gamma-rays were detected \citep{soderberg10_09bb,margutti14_12ap}.}
\tablenotetext{e}{For SN~2009nz/GRB~091117, \citet{berger11} did not provide us their published spectrum, so we had to digitize their Figure 1, which shows their smoothed spectrum. Thus, some of our analysis which requires the raw spectrum, could not be performed.}.
\tablenotetext{f}{While \citet{cano14} suggest that SN~2013ez/GRB~130215A is a SN Ic, SNID finds good matches with both SNe Ic-bl (including SNe Ic-bl~2007ce and 2005nb ) and with SNe Ic. Thus, we call it a "SN Ic/Ic-bl", since SNID does not find an unambiguous type, and we do not include it when constructing mean spectra of SNe Ic-bl. Furthermore, the redshift of this SN/GRB is uncertain, as there is only a lower limit from metal absorption lines superimposed on the GRB afterglow spectrum as mentioned in the GCN by \citet{cucchiara13_13ez_redshift}.}
\tablenotetext{$\dagger$}{For consistency, we used the transformations of \citet{bianco14} to convert the dates of maximum in the $B$- or $R$-band into that in the $V$-band, as no $V$-band data have been provided.}
\tablenotetext{$^{\ast}$}{These SN-GRB harbored low-luminosity GRBs.}
\label{table_snIcblsample}
\end{deluxetable*}

We list our SN spectral sample and their references in Table~\ref{table_snIcsample} for normal SNe Ic, and in Table~ \ref{table_snIcblsample} for SNe Ic-bl with and without observed GRBs. We only include SNe with secure spectral IDs (see \citealt{modjaz14} and below for more information), whose spectra have been published before February 2015 and were made available to us by the authors, and whose date of maximum light has been measured. With those criteria, we have included \nSNIc\ SNe Ic and a total of \nSNIcbltot\ SNe Ic-bl, both with and without observed GRBs. 

Amongst the latter group, we have included \nSNGRB\ SNe Ic-bl with GRBs, sometimes referred to as SN-GRBs\footnote{We were not able to include one SN-GRB, since the authors did not provide us with their published spectrum: SN~2008hw/GRB~081007 \citep{jin13}. In addition, there are other SN-GRBs with spectra that have been reported in circulars, but are not included here as their spectra have not been published yet: GRB~111211A \citep{postigo12_sngrb}, SN~2012eb/GRB~120714B \citep{klose12_sngrb12eb},  SN~2013fu/GRB~130831A \citep{klose13_sngrb13fu}, PTF14bfu/GRB~140606B \citep{perley14_sngrb14bfu}, for which in the last stages of preparing this manuscript, the submitted paper by \citet{cano15_14bfu} with spectra appeared on arXiv after February 2015, and thus, is not included here.}, with published spectra, and denote them as ``SNe Ic-bl with GRBs" in plots. For the SN-GRBs, we further distinguish SNe Ic-bl accompanied by low-luminosity GRBs (LLGRBs) as opposed to those with high-luminosity GRBs (HLGRBs), where the luminosity refers to the isotropic gamma-ray luminosity. LLGRBs, in addition to their low luminosities ($10^{46}- 10^{48}$ \ergs ), appear to be more isotropic and have relatively soft high-energy spectra compared to HLGRBs (e.g., \citealt{campana06,hjorth13,nakar15} and references therein). There have been suggestions that the high-energy emission of the GRBs in these two groups may come from different mechanisms (e.g., \citealt{waxman07,bromberg11,nakar12,nakar15}). We note that when corrected for beaming angle, the intrinsic energies of LLGRBs and HLGRBs are similar (e.g., \citealt{mazzali14}), and that there are also some GRBs that appear to be transition objects (e.g., SN~2012bz/GRB~120422A, \citealt{schulze14}) with intermediate luminosities, which leads us to collect the gamma-ray luminosities of all the SN-GRBs in this sample \citep{mazzali14}. In Table~\ref{table_snIcblsample}, we mark the following LLGRBs: SN 1998bw/GRB 980425, SN 2003lw/GRB 031203, SN 2006aj/GRB 060218, SN 2010bh/GRB 100316D. We furthermore analyze their SN spectra to see if there are differences between the spectra| features of the SNe connected with these two kinds of GRBs, and whether there are any correlations between SN spectral properties and GRB energies (all GRB energies taken from the compilation in \citealt{mazzali14}, see references therein).

In addition, we included SNe~Ic-bl 2009bb \citep{soderberg10_09bb,pignata11} and 2012ap \citep{margutti14_12ap,milisavljevic15_12ap} as ``engine-driven" SNe
since they were suggested to be relativistic SNe based on their large and early radio emission, similar to nearby SN-GRBs, even if no gamma rays were detected \citep{soderberg10_09bb,chakraborti15}. Thus, in our analysis below, we include SNe~2009bb and 2012ap in our sample of SN-GRBs. 

We also included SNe Ic-bl without detected gamma-ray emission and designate them as ``SNe Ic-bl without observed GRB" in the subsequent analysis, while referring to them simply as ``SNe Ic-bl" in the plots (as opposed to ``SNe Ic-bl + GRBs"). The presence of an off-axis or weak jet can be excluded for some, but not all, SNe Ic-bl without observed GRBs in our sample. For example, while for SN Ic-bl~2002ap, the radio observations were deep enough to exclude the presence of a jet, be it on- or off-axis \citep{berger02}, for SN~2010ay, \citet{sanders12_10ay} cannot rule out a low-luminosity GRB like GRB060218 from their radio and X-ray upper limits. In general in all SNe Ic-bl without observed GRBs, except in the most nearby SN~1997ef, 2002ap and SN~2003jd, the various X-ray and radio limits, when they exist (for PTF10bzf: \citealt{corsi11}; PTF10vgv: \citealt{corsi12}), only rule out very energetic relativistic GRB explosions, such as \sndh, but not necessarily low-luminosity GRBs such as \snbw\ or \snaj . With the caveat that the presence of off-axis and low-luminosity GRBs cannot be ruled out for all SNe Ic-bl in our sample, we proceed in the rest of the paper to designate these as SNe Ic-bl without GRBs, but discuss this caveat in more detail later on.


Unless found by the accompanying GRB trigger, the SNe in our sample were discovered by both galaxy-targeted SN surveys (e.g., LOSS, \citealt{filippenko01}) and galaxy-untargetted ones (e.g., Texas SN Search\footnote{\url{http://grad40.as.utexas.edu/~quimby/tss/index.html}.}, \citealt{quimby06_thesis}; SDSS II SN survey\footnote{\url{http://www.sdss.org/supernova/aboutsprnova.html}.}, \citealt{frieman08}). Here, we report the median redshifts for each of the SN subtypes in our sample: $z_\mathrm{Ic}$=0.0126, $z_\mathrm{Ic-bl}$=0.0232 for the SN Ic-bl without observed GRBs, and $z_\mathrm{SNGRB}$=0.1450 for the SN-GRBs. The fact that the different SN subtypes span somewhat different redshift ranges (with SN-GRBs at the highest redshifts, given their rarity) should not influence our results, as all SN spectra are de-redshifted into the rest frame, and all spectral comparisons are made for the spectral features that were observed in all SN spectra at all redshifts. In addition, any evolutionary effect is expected to be insignificant within such a small redshift range, for which e.g. the metallicity content of the universe does not change significantly \citep{tremonti04}.

We note that we did not include a new and very rare kind of SNe Ic - the Superluminous SNe Ic, which are up to 100 times more luminous than normal core-collapse SNe and for which the powering source is highly debated (e.g., \citealt{quimby11,chomiuk11}, see review by \citealt{gal-yam12}). They are not included here, since it is not yet clear how they are physically related to the ``normal" SNe Ic and SNe Ic-bl in our sample (see discussion in \citealt{gal-yam12}). There is a recent claim of a spectrum that looks like that of a super luminous SN Ic has been observed in connection with an ultra-long GRB \citep{greiner15}, though the spectrum is very noisy.

\subsection{SN spectra and their spectral types}\label{spectra_subsec}

The spectra were drawn from public archives (SUSPECT, now subsumed by WISeREP\footnote{\url{http://www.weizmann.ac.il/astrophysics/wiserep/}};\citealt{wiserep12}), as well as the CfA Supernova Archive\footnote{\url{http://www.cfa.harvard.edu/supernova/SNarchive.html}}, with a large fraction of SNe Ic published in \citetalias{modjaz14}, or we directly requested them from the authors in the listed references. In order to concentrate on the SN features alone, we removed superimposed emission lines due to host HII regions. We also removed any telluric absorption in the non-Berkeley and non-CfA spectra if it was relatively narrow and a simple interpolation could reasonably remove it (for more details on pre-processing the SN spectra, see \citealt{liu15}).\footnote{However, we highly encourage SN observers to remove telluric absorption in their spectra as part of their reduction process, using standard procedures \citep{matheson00_93j}.}

We also include all relevant SNe that had been initially included in the SNID release version 5.0\footnote{\url{http://people.lam.fr/blondin.stephane/software/snid/}} by \citet{blondin07}, though in some cases with a modified SN identification (see extensive discussion in \citetalias{modjaz14}).  We also performed our own spectral classification on a few of the SNe using our expanded set of SNID templates \citep{modjaz14,liu14} and found two cases in which the ID was different from that announced by the authors. 

For PTF 10vgv  \citep{corsi12}, we find a better match with SNe Ic-bl than with SNe Ic, with SNID finding very good matches with a number of SNe Ic-bl  (SNe 1997ef, 2006aj, 2003lw, 2007ru). While \citet{corsi12} classify it a SN Ic based on its measured \SiSix\ velocities (which we also measure to be low, see below), its optical spectra are clearly those of a SNe Ic-bl, as also mentioned in the initial discovery ATel by PTF \citep{corsi10atel}. Thus, following our SN classification approach outlined in M14, we reclassify PTF10vgv as a SN Ic-bl. For SN~2013ez/GRB~130215A, \citet{cano14} suggest that is a SN Ic based on their spectrum taken 16.1 days (rest-frame) after explosion, which corresponds to around $V$-band maximum light, though the exact date of maximum is not known. We checked this ID by using SNID with the afore-mentioned augmented set of SNID templates. With a phase limit of only spectra with phases less than 10 days after maximum, SNID finds good matches with both SNe Ic-bl (SNe 2007ce and 1997ef) and with SNe Ic (1990aa, 2004aw) - though none can fit the feature at around 8200 \AA\ well (observed wavelength). Thus, we call SN~2013ez/GRB~130215A a "SN Ic/Ic-bl" in the table, since SNID does not find an unambiguous type, and we do not include it when generating mean spectra (see below). In addition, since the redshift of this SN/GRB is uncertain and there is only a lower limit of $z$=0.597 from metal absorption lines superimposed on the GRB afterglow spectrum \citep{cucchiara13_13ez_redshift}, the derived velocities based on the various absorption lines will be systematically uncertain and thus are not included in the velocity analysis below.

For SNe 2004dn, 2005kl and 2007cl, which have either very early or late data where He lines may be too strongly visible, we also checked by comparing with mean spectra of SNe Ib and of SNe Ic for those epochs \citep{liu15} and found that they were all clearly not consistent with SNe Ib, but rather with SNe Ic. Thus we can call them bona-fide SNe Ic\footnote{In \citealt{modjaz14} we noted that for these SNe we could not exclude the possibility that they may have developed or had He lines over the phases not covered by our data. However, with the average spectra of SNe Ib and SNe Ic we constructed in \citet{liu15}, we can exclude that they are SNe Ib.}.

In total, we have included  \nspecSNGRB\  spectra of  \nSNGRB\  SN-GRBs, as well as \nspecSNIcbl\  spectra of \nSNIcbl\ SN Ic-bl without GRBs and \nspecSNIc\ spectra of  \nSNIc\ SN Ic, from the literature and as part of the Stripped SN data release \defcitealias{modjaz14}{M14}  \citep{modjaz14}. Out of those \nspectotaltotal\ spectra,  \nspeclatetime\ were taken later than 90 days after $V$-band maximum and thus were taken during the nebular phase, where the radiative transfer physics is different than during the photospheric phase. Thus, they are not analyzed here, but are listed for completeness.

\subsection{Determining the date of maximum light in an uniform fashion }\label{spectra_subsec}

Here, we describe our approach for ensuring that we consistently use the date of maximum light in the same rest-frame filter as our reference point for all SNe in our sample. Following \citet{blondin07}, we choose the $V$-band since most current SN photometry data sets have best-sampled light curves in the $V$-band (e.g., \citealt{bianco14}). For SNe, where the published literature had determined their date of maximum light in a filter other than the $V$-band, but published their $V$-band data (i.e., SNe 2004ge, 2011bm), we have determined the date of maximum in the $V$-band, using the Monte Carlo polynomial fitting method of \citet{bianco14}. For high-redshift objects (such as a few of the SN Ic-bl connected with GRBs), we also ensured that the filter we used corresponded to rest frame $V$-band filter. For objects where only $R$-band or $B$-band data were available (marked with a $\dagger$ in the table, so 8 out of 34 objects), we used the transformations of \citet{bianco14} (namely [Date of $V$-band max] = [Date of $R$-band max] - 1.8 days and [Date of $V$-band max] = [Date of $B$-band max] + 2.3 days, respectively, both with a standard deviation of 1.3 days, see their Table 10) to convert from the $R$- or $B$-band date of maximum, respectively, to that in the $V$-band. We note that for SN 2011bm, the date of maximum in the $V$-band is different by 11 days from that in the $R$-band - which is much larger than found for a large set of Stripped SNe in \citet{bianco14}. However, SN~2011bm appears to be an odd SN Ic, as it exhibits large ejecta masses (7$-$17 \sm), whereas most SNe Ic have much smaller ejecta masses, around 2-4 \sm\ (e.g, \citealt{drout11,cano13,lyman14}, Bianco et al., in prep).
In addition, for SN 2004fe, there were multiple datasets in the literature, and we used the date of $V$-band maximum as measured by \citet{bianco14}, which is different from that in \citet{drout11} by 1.3 days.


\section{Quantifying SN diversity by constructing Average and Standard Deviation Spectra}\label{averagespec_sec}

For subsequent analysis, it is important that we remove the effect of any host galaxy reddening on the spectra we compare, and that we find a robust way to compare spectra taken with a heterogeneous set of telescopes (with different binning and wavelength ranges). To that aim we follow the same steps for pre-processing all spectra as in \citet{liu15}(see details in their Section~3.1), who in turn follow \citet{blondin07}. While the primary purpose of the spectra in \citet{blondin07} is to be included in the spectral library of SNID, a code for SN classification, here we use them to characterize the different SN classes. In summary: we binned the spectra to a common logarithmic wavelength scale and in that way mapped them onto the common wavelength range of 2500$-$10000 \AA. We then removed the shape of the pseudo-continuum, in order to focus on the characteristic absorption lines, and to not have to worry about reddening effects on the continuum (redding is investigated separately in photometry papers, e.g., \citealt{drout11}), or about any potential GRB afterglow light (which would be a power law continuum) in the GRB-SN spectra (e.g., \citealt{matheson03,deng05}). In addition, we checked that the default spline from SNID was reasonable even for the broad spectral features of broad-lined SNe Ic. Thus, the rest of the analysis and comparisons between SNe Ic and SNe Ic-bl should be robust as they are based on spectra flattened with the same number of spline knots. 

We have already released the constructed set of SNID spectral templates based on the \citetalias{modjaz14} data in \citet{liu14}, which are publicly accessible via \url{http://www.cosmo.nyu.edu/SNYU/} -  the newly constructed set of SNID spectral templates based on literature data, as presented here, are released on our SNYU github page (\url{https://github.com/nyusngroup/SESNtemple}), along with the data products discussed below (in particular, the mean and standard deviation spectra).

We note that the maverick PTF12gzk  displays very high Doppler absorption velocities (\citealt{benami12} and see our Fig.~\ref{Vel_comp_fig}) similar to those of SN Ic-bl  (and also based on radio observations \citealt{horesh13}), but does not display broad lines. Thus, we do not include PTF12gzk when constructing mean spectra (see Section~\ref{averagespec_sec} for details).

\subsection{What is the ``typical SN Ic" or ``typical SN Ic-bl" ?}\label{average_subsec}

In a number of crucial situations, one frequently encounters the question ``What does a typical SN Ic look like?" or ``Is there a typical SN Ic-bl ?" Such situations include when intending to determine the type of a newly discovered SN, when trying to determine if a new transient is a novel type of explosion (e.g., \citealt{kasliwal12}) or when modeling the spectral contamination of SNe Ic in SNe Ia searches (e.g., \citealt{poznanski02,graur13,graur15}). Historically, nearby and therefore well-observed SNe have been taken as the best and representative example, as in the case of SN 1994I (e.g.,
\citealt{filippenko95,richmond96}), which was a SN Ic in the nearby M51, or \snbw\ \citep{galama98} referred to as the ``typical" SN-GRB, which was the first, and closest, instance of the SN-GRB connection. 

\begin{figure*}[!ht]                
\vspace{-1.5in}
\hspace{.7in}
\centerline{
\includegraphics[width=1.4\columnwidth,angle=0]{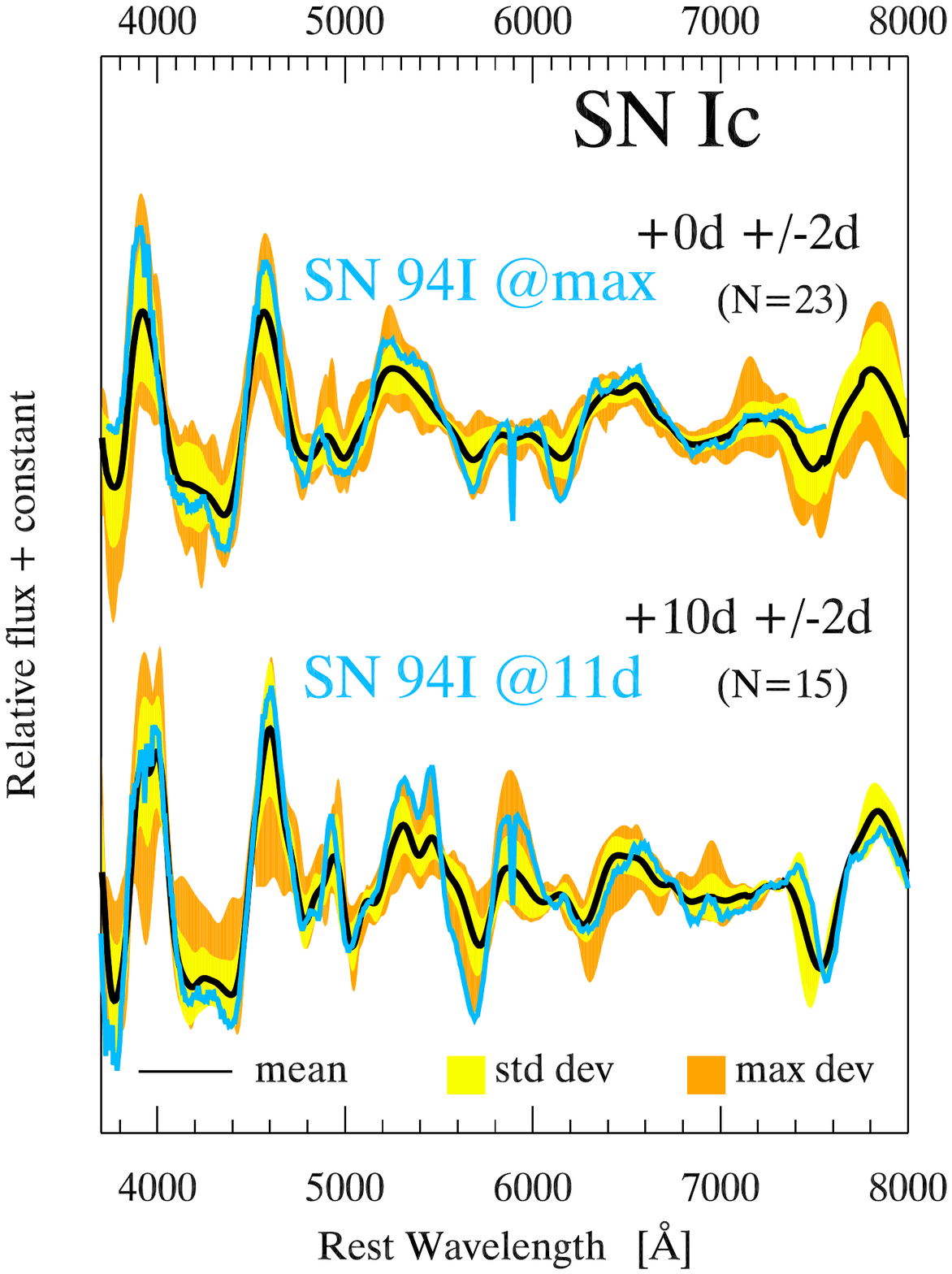}
\hspace{-1.in}
\includegraphics[width=1.4\columnwidth,angle=0]{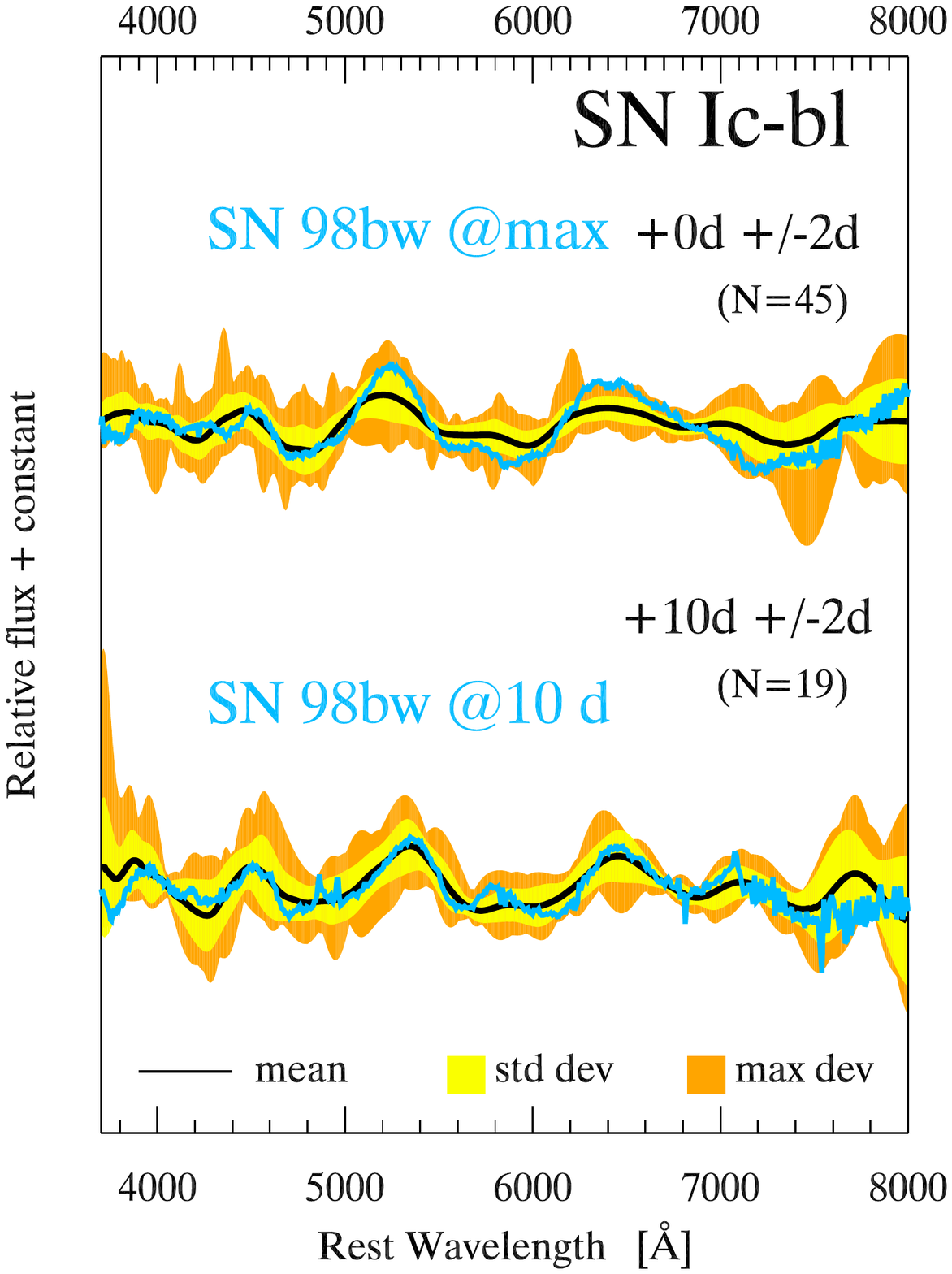}
}
\singlespace \caption{Mean spectra of SNe Ic (in black, left) and of all SNe Ic-bl (including both with and without GRBs, right)  with corresponding standard deviation (in yellow) and maximum deviation (in orange), after we removed the pseudo continuum (see text for more details), at peak brightness $\pm$ 2 days (top), and at 10 days after peak $\pm$ 2 days (bottom) in the rest frame. SNe 2013ez and PTF12gzk are not included here (see text).
Shown are also the number of individual spectra that were used to construct the mean spectrum for each epoch range. For comparison, overlaid is the SNID-ified spectrum of SN~1994I (left, \citealt{filippenko95,modjaz14}) and of SN~1998bw (right, \citealt{galama98}), which have been considered as the standard SN Ic and SN Ic-bl, respectively. The spectra of SN 1994I are more than one standard deviation away from the mean spectra of SNe Ic, often reaching maximum deviation, at many wavelengths. On the other hand, SN~1998bw appears to be a typical SN Ic-bl - its spectrum is very close to the mean spectrum and well within one standard deviation, specially at 10 days after max. We note that there were twice as many SN-GRBs than SN Ic-bl without observed GRBs that went into constructing the SN Ic-bl mean spectrum, and thus, that our mean SN Ic-bl spectrum is weighted more towards SN-GRBs spectra. Note that the apparent line at 7400$-$7800\AA\ and its large standard deviation in the mean spectra is probably due to a combination of \OxySeven\ and telluric (i.e., atmospheric) absorption (A-band at 7620 \AA ); the latter was removed only for the \citetalias{modjaz14} and Berkeley spectra.}
\label{SNIcIcblmean_fig}
\end{figure*}                                                                 

However, the natural next step, given the large amount of data we have, is to employ a statistical approach, by computing the mean and standard deviation spectra of a specific type of SNe at a specific phase, to investigate the amount of homogeneity, or the lack thereof, in \emph{all} SNe of the same class. We have done so for the spectra of SNe Ic and SNe Ic-bl (this work) and for the spectra of SNe Ib and SNe IIb \citep{liu15}, the four general categories of stripped SNe .

We construct the mean spectrum, as well as the standard and maximum deviation from the mean, following \citet{blondin12} and \citet{liu15}, for SNe Ic and for SNe Ic-bl.
The results are shown in Fig.~\ref{SNIcIcblmean_fig} for two specific phases: at maximum light and at 10 days after maximum light (for mean spectra at other phases, see Appendix), where all phases are in the rest-frame of the objects. For all cases, we included spectra that are within $\pm$2 days of those dates. On Fig~\ref{SNIcIcblmean_fig} as well as on subsequent figures, we also note the number of individual spectra that were used to construct the mean spectrum for each epoch range. Sometimes, there is more than one spectrum per SN included (especially for mean spectra at maximum light) as we included multiple spectra of the same SN from various groups with different wavelength ranges, as well as multiple spectra of the same SN obtained over the four-day bin. In order to check that the mean spectra are not dominated by a few SN with large numbers of spectra, we also constructed mean spectra, where we include only one spectrum of each individual SN. There is no statistical difference between mean spectra based on one spectrum per SN vs. those based on multiple spectra per SN.  We note that the mean spectrum of broad-lined SNe Ic include more spectra of SN-GRBs than those of SN Ic-bl without observed GRBs (see Section~\ref{snlc_blwithandwithoutGRBs_subsec}), such that it is weighted more towards SN-GRBs spectra. This is due to the apparent higher interest in observing and following up SN-GRBs than SN Ic-bl without observed GRBs.

In addition, we test whether the so-called ``prototypical" SN Ic 1994I and SN Ic-bl SN~1998bw are truly representative of the respective SN sub-types by comparing them to the mean spectra for that class. Fig. \ref{SNIcIcblmean_fig} shows that SN 1994I is more than one standard deviation away from the mean SN Ic spectrum, even reaching the maximum deviation at many wavelengths, meaning its spectra are the most deviant of the SN spectra in the sample - at both maximum light and 10 days after max. Thus, we conclude that SN~1994I cannot be called a typical SN Ic spectrally, in contrast to common belief. The comparison of its photometric properties reveals a similar and even more pronounced conclusion, namely that it is not representative of light curves of SNe Ic (\citealt{richardson06,drout11}, F. Bianco et al, in prep.), since it faded much faster than the average SN Ic light curve. On the other hand, SN~1998bw seems to be a ``typical" SN Ic-bl (Fig.~\ref{SNIcIcblmean_fig}) -  at both phases, SN~1998bw  is consistent with the mean spectrum well within one standard deviation at most wavelengths for its spectrum at $t=+$0 days, and at most wavelengths for its spectrum at $t=+$10 days.  While we only show mean spectra at two specific epochs, our conclusions remain unchanged when making this comparison at other phases. Thus, using SN~1998bw as a spectral template for other broad-lined SN Ic and SN-GRBs (e.g., most recently \citealt{cano14}) appears warranted, but note that our SN Ic-bl mean spectrum is more weighted towards spectra of SN-GRBs.

\subsection{The Helium Problem for SN-GRBs - is there Helium hidden in SN Ic-bl?}\label{he_subsec}

From the massive stellar evolutionary point of view, it is very hard to produce a SN-GRB progenitor which is both H- and He-free at the low metallicites at which most SN-GRBs are observed \citep{modjaz08_Z,levesque10_grbhosts,graham13}. Indeed, most of the models for GRB progenitors are He-stars \citep{woosley02,yoon05,woosley06_z,yoon15}, with He masses between 1.0 $- $ 15 \msol\ in the case of single low-metallicity progenitor stars with Zero-Age-Main-Sequence masses of 15 $- $ 50 \msol\ \citep{woosley02}. While hiding He due to insufficient mixing has been considered \citep{hachinger12,dessart12,piro14,liu15}, the amount that can be hidden appears to be very small ($<$0.1 \msol\ of He; \citealt{baron96,hachinger12}), much smaller than the expected mass of the He progenitor star. Another possibility may be that the He lines are present but not detectable in SNe Ic-bl spectra due to smearing by the high ejecta velocities in SNe Ic-bl. The latter is what we want to address here by simulating the smearing due to high velocity on real SNe Ib spectra, i.e. those with detected He lines - effectively simulating a ``broad lined SN Ib". We then want to explore if those broadened He lines could have been detected in spectra of SNe Ic-bl and SN-GRBs.

\begin{figure}[!ht]    
\vspace{-1.5in}
\hspace{-.2in}
\includegraphics[scale=0.6,angle=0]{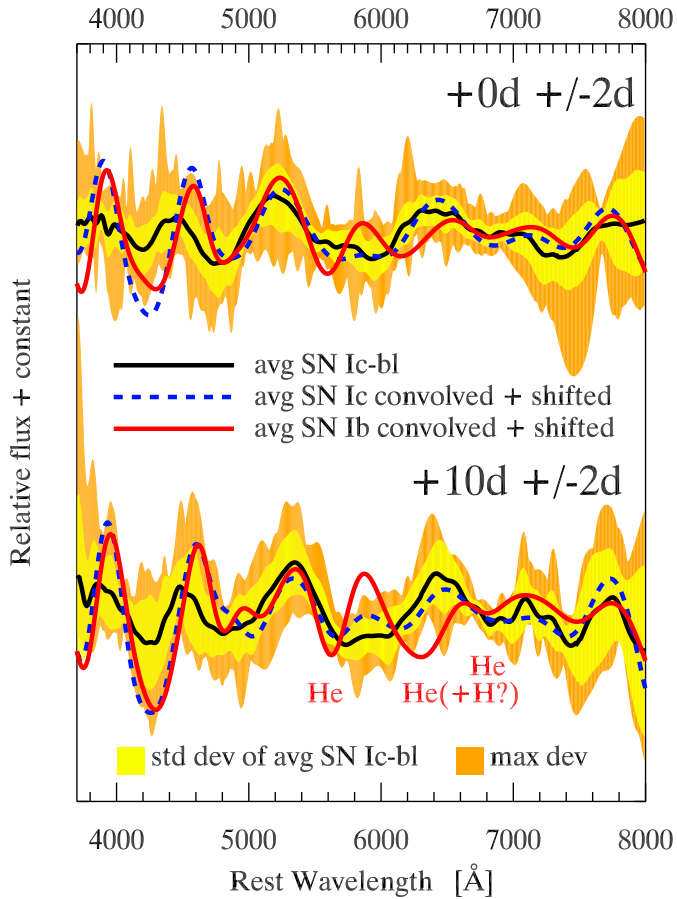}
 \caption{Mean SN Ic-bl spectra for two different phases (in black), as well as an average SN Ic and an average SN Ib spectrum - the latter two convolved with a Gaussian and blue-shifted. The average SN Ic-bl spectrum can be reasonably reproduced by a normal SN Ic spectrum convolved with a Gaussian with $\sigma$=6,000 \kms Gaussian (i.e., FWHM=14,000 \kms), and blueshifted by 3,000 \kms. However, a similarly convolved SN Ib spectrum deviates from the average SN Ic-bl spectrum at the positions where the He-lines are expected, and is well outside the standard and maximum deviations (the latter not shown) - indicating that even a broad-lined SN Ib, i.e., a smeared out SN Ib spectrum, would have had detectable signs of He lines, which are not observed in our observed SN Ic-bl mean spectra. Annoted are the lines of \HeFive , \HeSix\ (which may be blended with H$\alpha$, as SNe Ib may show some H$\alpha$; \citealt{parrent16}, \citealt{liu15}), and \HeSeven\ in the spectrum of the simulated "SN Ib-bl". }
\label{meanspec_conv}
\end{figure}    

For this purpose, we take the SN Ic mean spectra from Section~\ref{average_subsec}, as well as the SN Ib mean spectra\footnote{The SN Ib mean spectrum for maximum light is based on 38 spectra of 13 SNe Ib, while the SN Ib mean spectrum for 10 days after maximum is based on 37 spectra of 18 SNe Ib.} (constructed in the same way, see \citealt{liu15}) and convolve both with a Gaussian of width $\sigma$=6,000 \kms (i.e., FWHM=14,000 \kms), choosing a conservative value, based on the results of our template fitting method discussed below (see Fig.~\ref{SNIcbl_sigmaCDF_fig}). We note that in order to match the spectral features of the mean SN Ic-bl spectrum, in addition, we blue-shifted both the convolved mean SN Ib and the mean SN Ic spectra by 3,000 \kms. A convolution with a Gaussian profile (or inverse Gaussian smoothing) simulates well the smearing of the spectra due to increase of kinetic energy per unit mass (\citealt{iwamoto98}; K. Nomoto, private communication). The results are shown in Fig.~\ref{meanspec_conv}: while convolving a SN Ic mean spectrum with a Gaussian profile yields a spectrum that is very similar to a SN Ic-bl mean spectrum (except over a small wavelength range of 4000$-$5000 \AA, most likely due to Fe blending, see below), a similarly convolved SN Ib mean spectrum is different from the observed SN Ic-bl mean spectrum, by more than 3-4 standard deviations at the positions where the He lines are expected, especially for spectra at 10 days after maximum. The smeared-out He lines in the convolved SN Ib spectrum are even outside the maximum deviation spectra of the observed SN Ic-bl mean spectrum,  especially for spectra days after maximum light. The match between the SN Ic-bl and our smeared SN Ic spectrum is not unique; a small range of convolution kernels and blueshifts provide similarly good fits. In all such cases, using the same fit parameters for simulating a ``broad-lined SN Ib", the broadened He-lines should have been detected in spectra of observed SNe Ic-bl, including those of the SN-GRBs, but they are not. Indeed, the choice for the convolution parameter shown in Fig.~\ref{meanspec_conv} is more than conservative, since it is at the high end of the range of convolution parameters when doing a formal template fitting procedure for the \FeFive\ line, meaning it is amongst the most extreme, i.e. worst, cases of smearing-out for SNe Ic-bl and SN-GRBs (see Fig.~\ref{SNIcbl_sigmaCDF_fig}).

Our conclusions remain unchanged when performing the same comparisons for mean spectra at different phases, as long as they are after maximum light, when the He signature is strong enough in SNe Ib for a robust detection \citep{liu15}. In the same spirit, we repeated this exercise by simply adding He lines into the SN Ic mean spectra. Again, we find that smeared-out He lines do not match the appropriate region in the SN Ic-bl spectrum, even when using very high smearing velocities (up to 25,000 \kms\ with respect to the SN Ic spectrum). Thus, we conclude that even a smeared-out SN Ib spectrum (i.e., a simulated SN ``Ib-bl"),  would have had detectable He lines, which are not observed in spectra of SNe Ic-bl and SN-GRBs, nor is it possible to smear out the He lines sufficiently make them non-detectable. The lack of He lines in SN Ic-bl-GRB  spectra is in line with findings by e.g., \citet{taubenberger06} and \citet{chornock10}, who obtained NIR spectra of SNe Ic-bl and did not observe the strong 1\micron\ line of He. Recently, \citet{milisavljevic15} claimed a potential detection of the 1 \micron\ line of He in SN Ic-bl 2012ap via SYNOW. However, since they do not detect the otherwise clean and strong 2 \micron\ line of Helium in the same spectra at phases where one expects it, their evidence for significant He detection is weak and the 1\micron\ line feature may be due to other lines, as had been suggested before for other SNe \citep{millard99,gerardy04,sauer06,friesen14}.

\begin{figure*}[!ht]    
\vskip-1.cm 
\vspace{-1.5in}
\hspace{.7in}
\centerline{
\includegraphics[width=1.4\columnwidth,angle=0]{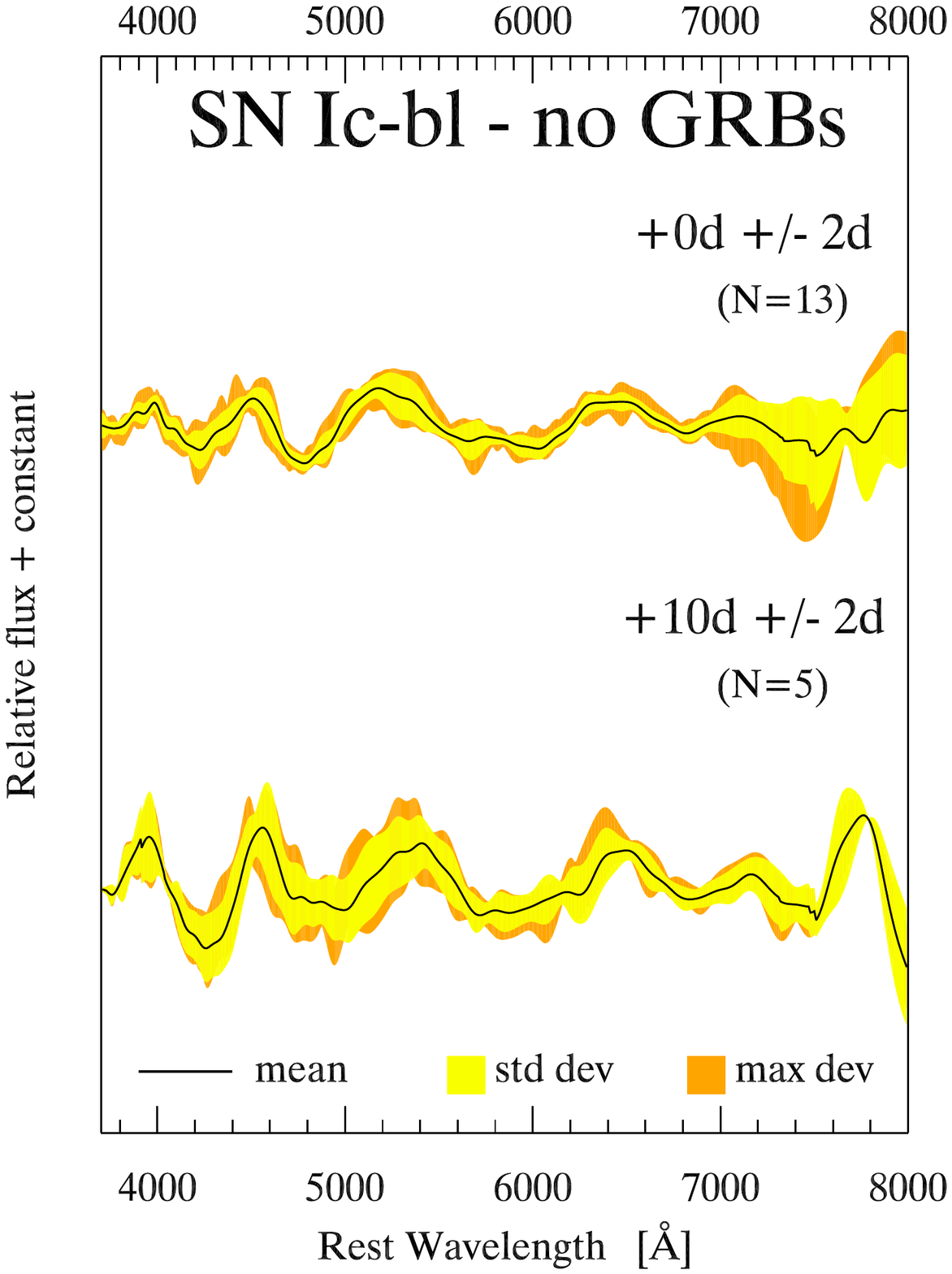}
\hspace{-1.1in}
\includegraphics[width=1.4\columnwidth,angle=0]{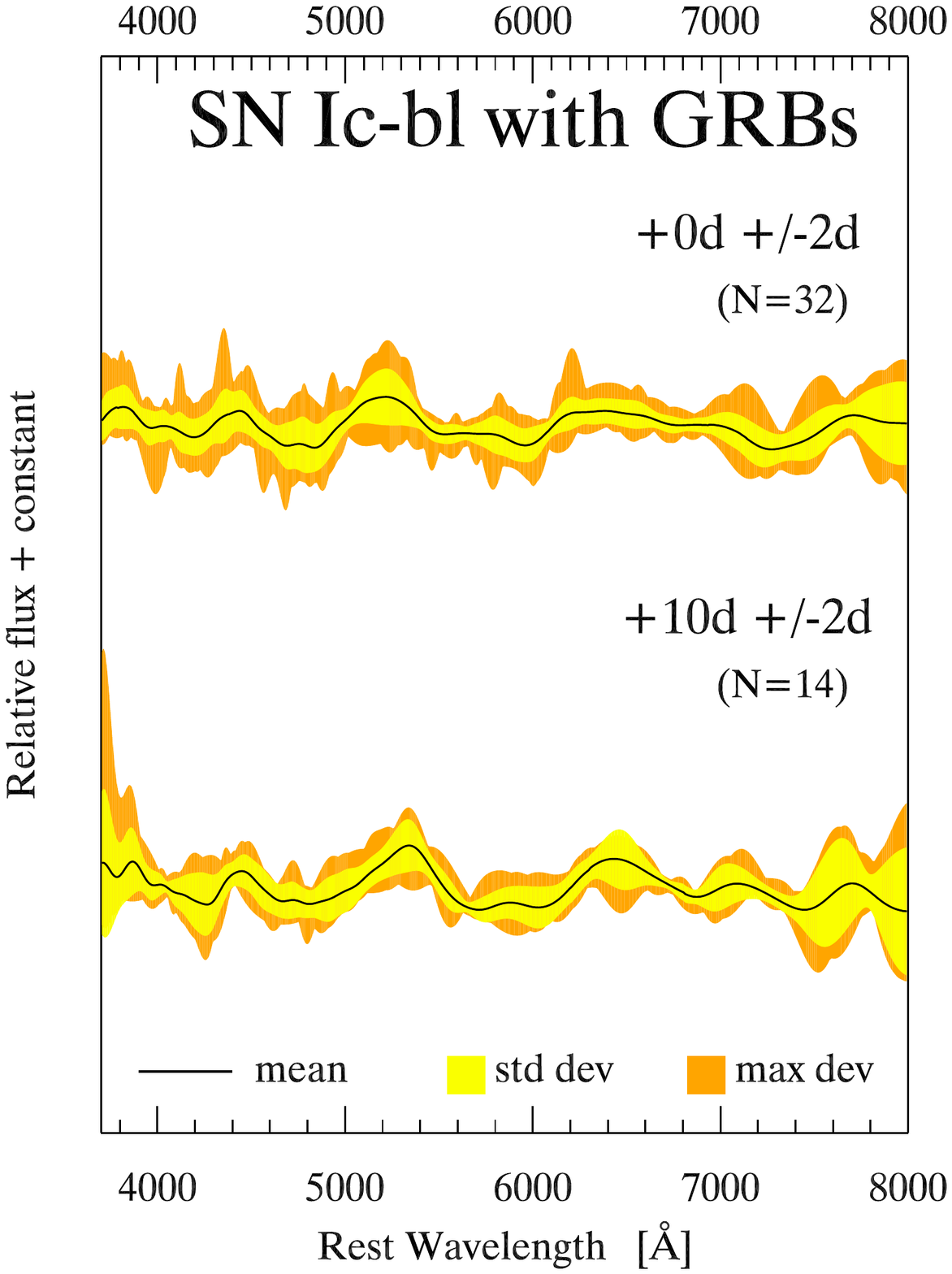}
}

 \caption{As in Fig.~\ref{SNIcIcblmean_fig}, but for average spectra of SN Ic-bl without observed GRBs (left) and SN Ic-bl with GRBs (right), for two different phases. SN-GRB have broader lines than SN Ic-bl without GRBs (see text for more details). }
\label{meanspec_SNIcbl_withwithoutGRB}
\end{figure*}    

Thus, we conclude that the absence of clear He lines in optical spectra of SNe Ic-bl is not due to the smearing out of the Helium lines by the high velocities present, as otherwise they would have been still detectable, assuming that the same process that operates on all the other lines affects the Helium lines in the same way. We discuss the implications these findings have for the understanding of SN-GRB progenitor systems in Section~\ref{discussion_sec}.
Nevertheless, we note that there is a large discrepancy between the simulated SN Ic-bl spectrum (i.e., after broadening the SN Ic mean spectrum) and the real SN Ic-bl mean spectrum in the blue region (3500-4400 \AA, Fig.~\ref{meanspec_conv}), where the forest of Fe lines lie, which may have implications for understanding line formation in highly expanding atmospheres. We encourage sophisticated spectral synthesis models such as CFMGEN \citep{dessart07}, PHOENIX \citep{baron95}, and SEDONA \citep{kasen06} to model SNe Ic-bl spectra, which has not been done before.

\subsection{Differences in Spectra between SNe Ic-bl with and without GRBs}\label{snlc_blwithandwithoutGRBs_subsec}

Two of the outstanding questions in the field are what distinguishes SNe Ic-bl \textbf{with} observed GRBs from those \textbf{without} observed GRBs,  and what produces SN Ic-bl, including those without observed GRBs. While host galaxy analysis have shown that there may be differences in metallicity and SFR density \citep{modjaz08_Z,sanders12,kelly14}, here we study the explosion properties, and in particular the early-time, photospheric spectra. To that aim, we construct mean spectra of SNe Ic-bl \textbf{without} observed GRBs and mean spectra of SNe Ic-bl \textbf{with} GRBs and plot them for two epochs (around maximum light and 10 days after) in Fig.~\ref{meanspec_conv} side-by-side. We include SN~2009bb amongst the SN-GRBs, since it was suggested to be a relativistic SN based on its large radio-emission, similar to nearby SN-GRBs, even if no gamma-rays were detected \citep{soderberg10_09bb}.

In Figure~\ref{meanspec_SNIcbl_withwithoutGRB}, we can see that the features in SNe Ic-bl with GRBs are somewhat broader than those in spectra of SNe Ic-bl without GRBs, especially at 10 days after $V$-band maximum. For example, for the SNe Ic-bl without GRBs at $t_{Vmax}=+10$ days, the wavelength range between 4000\AA\ and 5000 \AA\ appears to have a number of "features" $-$ 2-3 troughs $-$, while the same wavelength range for the SN-GRBs shows only very smooth undulations and at most one broad feature. We note that our conclusions remain unchanged even if we restrict ourselves to include only one spectrum per included SN, in order to avoid that many spectra of the same SN dominate the mean spectra. As shown below in Section~\ref{Fevabs_subsec}, this trend is also seen quantitatively in the amount by which the \FeFive\ line in the mean spectrum of SNe Ic has to be broadened in order to match the SN Ic-bl spectra (i.e. the convolution velocity of the template fitting method, see also Appendix Section~\ref{speclineID_sec}). We will address this question using a complementary method, namely measuring the Doppler-shifted absorption velocity of \FeFive\ in Section~\ref{specline_sec}, where we find the same trend, but for Doppler-shifted absorption velocities: SN Ic-bl/GRB have higher velocities, on average, than SNe Ic-bl without observed GRBs. We will discuss the implications of these observations for the SN-GRB explosion mechanisms and production mechanisms of GRBs in Section~\ref{discussion_sec}.

\section{Spectral Line Velocity Evolution}\label{specline_sec}

Now we shall shift our attention to measuring velocities, since they provide clues about the dynamics of the explosion, and since an estimate of the photospheric velocity is needed for computing important quantities, such as ejecta masses, from light curves and spectra (e.g., \citealt{drout11,cano13,lyman14,taddia15}), using analytical models (e.g., \citet{arnett82}.  Here, it is crucial to distinguish between two main kinds of velocities: the absorption velocities, called $v_{abs}$, which are measured from the (Doppler) blue-shifted trough of specific lines, and the width velocities, by which the lines are broadened (which we call $v_{convolution}$ based on the method we use, see Appendix~\ref{speclineID_sec}). We measure both separately $-$  show in Section~\ref{specline_94Isubsec} that different lines and different methods give different absorption velocities $-$ and present their self-consistent analysis in Section~\ref{Fevabs_subsec} for \FeFive\ absorption velocities, and in Section~\ref{Fewidths_sec} for the width velocities  respectively. As we will see below, while for most cases these two kinds of velocities correlate with each other and are very similar, as expected from radiative transfer models in the kinds of expanding atmospheres considered here (e.g., \citealt{dessart11}), there are outliers: cases where a SN has high absorption line velocities, but small line widths (e.g., PTF~12gzk) and vice versa, a SN that has large line widths but low absorption velocities (e.g., PTF~10vgv). Thus, until a unifying picture of spectral line formation in SN Ic and SN Ic-bl is developed and tested (e.g., including asphericities in the ejecta geometry), we suggest that authors, when reporting velocities, are clear about how the number was derived (whether from modeling or observations, and which line was used) and whether it is the absorption velocity or the width of the line that they are reporting.

\subsection{Absorption line velocities and their measurement errors}\label{smoothing_subsec}

We determine the blueshifts in the troughs of the absorption lines from the \citetalias{modjaz14} Stripped SNe sample, as well as those from the literature (see Table 1). For the SNe Ic spectra, we use the robust technique of measuring SN line profiles and of estimating error bars as presented by us in \citet{liu15}. Here we summarize the salient features: we constructed noise-smoothed spectra and measured the location of maximum absorption using the procedure recommended by \citet{blondin12} of fitting a parabola to a small region around the minimum absorption. It is especially powerful since it does not assume specific, Gaussian, for the full line profile and therefore allows for asymmetric shapes for the full line profile. We refer to our velocity measurements as Òabsorption line velocitiesÓ (in short "$v_{abs}$") in the rest of this work. We verified that our velocity measurements are robust: they yielded the same values as on the continuum-removed spectra, which were produced during the construction of mean spectra. As we showed in \citet{liu15}, we constructed spectral uncertainty arrays to be used for the subsequent estimation of the velocity measurements errors. We constructed spectral uncertainty arrays from the actual spectra themselves by separating noise from signal in Fourier space, taking advantage of the fact that the typical width of a supernova spectral feature is large, since it is Doppler broadened by the $\sim$ 5,000 - 20,000 \kms\ expansion velocity of the SN ejecta. For computing the errors on the line identification velocity measurements, we conducted Monte Carlo simulations where we generated 3,000 mock spectra, using the associated spectral uncertainty arrays, and reran our automated velocity measurement code on them (for more details see \citealt{liu15}).



We did this analysis for \FeFive\ , since  \FeFive\ has been suggested to trace well the photospheric velocity (\citealt{branch02}, but see \citealt{dessart15} for caveats). On the other hand, as we discuss below, while \SiSix\ is commonly identified with the most isolated feature at $\sim$6000$-$6100 \AA\ in broad-lined SNe Ic (e.g., \citealt{patat01,chornock10,lyman14}), the identification of \SiSix\ in both SNe Ic and SNe Ic-bl spectra may be problematic \citep{parrent16}. We note that our line identifications are not definitive since we have not conducted spectral synthesis calculations for all spectra. For clarity, in Appendix~\ref{speclineID_sec} we describe how we self-consistently identify \FeFive\ for all subtypes of SN Ic.

However, this procedure (which we call the "line identification method" hereafter) does not work properly for spectra of SNe Ic-bl, since those spectra are severely blended - in particular, the multiple lines of \ion{Fe}{2} (\FeFour\ , \FeFivezero\ , and \FeFive\ ), which are visible in SNe Ic, are blended into one broad feature in SNe Ic-bl (see Fig.~\ref{Fe-ID_fig}). As we detail in Appendix~\ref{speclineID_sec}, here we develop a novel method to measure line velocities and widths for SNe Ic-bl using a template-fitting method. In summary, we use the mean spectrum of SNe Ic that we constructed in Section~\ref{averagespec_sec}) (in particular the region around \FeFive ), blue-shift it and convolve it with a Gaussian (similar to what we did in Section~\ref{he_subsec}) to find the convolution width and blueshift that best matches the input SN Ic-bl feature. We implemented this process by using the Monte Carlo Markov Chain (MCMC) sampler "emcee" \citep{foreman-mackey13} to find the best fit and to obtain error bars. This procedure usually yielded very good fits as indicated by the $\sim$ unity values for the min $\chi^2$ values (e.g., see figure~\ref{fig_Icbl_Ic_fit}). We tested that both methods (the identification method and the template fitting MCMC method) yielded the same values of $v_{abs}$ for SNe Ic for which both methods can be applied. Thus, we think that there is no systematic shift or bias when comparing the $v_{abs}$ values of SNe Ic (obtained via the line identification method) to those of SNe Ic-bl (obtained via the convolution MCMC method). We compare our derived velocities to those from the literature in Section~\ref{litcomparison_subsec}. For the low S/N spectra of \snma, no satisfactory fits could be achieved, so we do not include any of the fitting results for \snma.

We emphasize that even if the feature we identify as \FeFive\ is not due to \FeFive , relative comparisons between the different SN types will still hold and are valuable, since we are self-consistently using the same feature for velocity measurements for \emph{all} SN Ic sub-types. As to \SiSix , there are some suggestions that the commonly identified feature may not be due to \SiSix, but rather a combination of elements, including H$\alpha$ \citep{parrent16}. Furthermore, there are sometimes two features (one at 6100 \AA\ and one at 6300 \AA) in that range, and without spectral synthesis calculations for all spectra it is hard to identify which one is due to \SiSix . While we have measured the absorption wavelength of the strong feature at $\sim$6300 \AA\ for SNe Ic, and for SN~1994I, we show absorption velocities based on the identification of that feature with \SiSix\ only for comparison purposes (Fig.~\ref{sn94Ivels_fig}) and do not report them for our full sample. In addition, for SNe Ic-bl we find from our template-fitting that the SN Ic template (after blue shifting and broadening) is not a good representation of the SN Ic-bl spectra in the wavelength range of $\sim$5800 \AA $-$6300 \AA\, and we do not obtain good fits (Appendix~\ref{speclineID_sec}).

Using the two different methods above, we list the measured velocities and their error bars for the \FeFive\ line as a function of spectral phase for all SNe in our sample in Table~\ref{vels_table}. This table is available in its entirety in a machine-readable form from the online journal. The listed and plotted error bars represent the 68\% uncertainty range.

\begin{deluxetable}{ll}
\tabletypesize{\scriptsize}
\tablecaption{Measured absorption velocities\label{vels_table}}
\tablehead{
\colhead{Phase\tablenotemark{a}} &
\colhead{$v_{abs}$(\FeFive )} \\
\colhead{(days)} &
\colhead{(\kms )} 
}
\startdata
\multicolumn{2}{c}{\bf sn1983V} \\
-15.0 &   $-$18000 $\pm$ 970 \\
-12.0 &  $-$19000 $\pm$ 1100 \\
-10.0 &   $-$19000 $\pm$ 880 \\
-2.0 &   $-$12000 $\pm$ 940 \\
5.9 &   $-$10000 $\pm$ 460 \\
6.9 &   $-$10000 $\pm$ 400 \\
7.9 &   $-$10000 $\pm$ 430 \\
8.9 &    $-$9900 $\pm$ 540 \\
9.9 &    $-$9500 $\pm$ 530 \\
10.9 &    $-$9300 $\pm$ 510 \\
11.9 &    $-$9100 $\pm$ 550 \\
31.8 &    $-$9000 $\pm$ 680 
\enddata

\tablecomments{This table is available in its entirety in a machine-readable form in the online journal. A portion is shown here for guidance regarding its form and content.}
\tablenotetext{a}{Rest-frame age of spectrum in days relative to $V$-band maximum. See text for details.}

\end{deluxetable}



\subsection{Comparison with literature values of Absorption velocities for SN-GRBs and Stripped SNe}\label{litcomparison_subsec}

Two recent papers have measured velocities in large samples of Stripped SN spectra that include SNe that are analyzed in this work. Since they did not compute velocity uncertainties, the following comparison will include only our velocity error values. 

\citet{schulze14} include the same seven SN-GRBs as we, mostly with the same published data. However, they attempt to measure \FeFive\ with a method that is different from our main method of template fitting, but somewhat similar to our line identification method: they attribute the blended feature at $\sim$ 5000 \AA\ to \FeFive\ alone. Of the seven SN-GRB, where we and \citet{schulze14} use the same spectra published by them and others, their values for five SN-GRBs are consistent with ours within our error bars\footnote{For SN~2003dh/GRB030329, we note that its phases are not correct (Fig. 5 of \citealt{schulze14}), as the phases were de-redshifted twice (G. Leloudas, private communication). Adopting the correct phases, their values are consistent with our measured values within the uncertainties.}. For SN 2013cq/GRB~130427A, where both velocities are formally consistent within the error bars, their value is larger by  $\sim6000$ \kms. Also, for SNe 1998bw and 2003lw, their velocity values are all systematically larger than ours (by 12,000 \kms\ on average, which corresponds to more than 3-$\sigma$ errors). This discrepancy is most likely due to the fact that their method assumes that the whole feature is only due to \FeFive , and they do not include any contributions from the other two Fe II lines (namely \ion{Fe}{2} $\lambda\lambda$ 4924, 5018), which are blended with \FeFive\ in many of the spectra of SNe Ic-bl  (see Fig.~\ref{Fe-ID_fig}). In addition, without attempting to give an interpretation, we note that for some SN Ic-bl spectra the relative strength of the two parts of the Fe~II "W" feature are the opposite compared to the SN Ic template spectrum. This is the case for SN 1998bw and SN 2002ap at some, but not at all, phases. However, this does not affect our velocity measurements. The template fitting is most sensitive to the width of the entire complex, and in all instances where the relative feature strength is opposite to that in the template, we find that the edges of the feature coincide with the edges of the stretched and shifted template, and we conclude that the broadening and location of the feature are properly measured. For SN~1998bw, in particular, we think that our method is superior to the identification method (e.g., \citealt{schulze14}), as the latter identifies the wrong feature with \FeFive , namely the one that is actually due to the blend of  \FeFour\ and \FeFivezero. We think that this mis-identification is due to the fact that, as mentioned above, for \snbw\ the feature due to  \ion{Fe}{2} $\lambda\lambda$ 4924, 5018 is stronger than the one due to \FeFive, which is the other way around compared to most other SNe, and that \citet{schulze14} presumably used the dominant feature in that complex to identify as \FeFive . We did try to perform a fit in which we forced the dominant feature at $\sim$4700\AA\ to be due to \FeFive\ but those fits were ruled out.


We have 14 SNe of Types Ic and Ic-bl in common with the SN sample of \citet{lyman14} who use mostly the same data sets as we do, but measure Fe II velocities for some SNe and \SiSix\ velocities for others (from spectra taken closest to maximum light) in a highly inhomogenous way. While we fit a cubic spline to the distinct absorption feature of \FeFive\ in SN Ic spectra and fit a SN Ic template (consisting of the SN Ic mean spectrum) to the blended absorption features of triple Fe II in SN Ic-bl spectra, they fit a simple Gaussian profile  to absorption features of Fe II or Si II. However, for some SNe, they take values from the literature (either from spectral synthesis or from interpolating the velocity values from spectra taken at epochs far from maximum light)  and for others, even use mean values within SN subtypes in cases they could not identify Fe or Si\footnote{We note that the total number of SNe for all types we have in common with \citet{lyman14} is 26 and the conclusions using the full data set remain the same.}. 
For seven of the SNe, the velocity values are consistent with ours within our error bars (ie. within 1 $\sigma$, while for four SNe we are consistent at 1$-$2 $\sigma$ and three SNe we are consistent within $\sim$2 $\sigma$. Given that the measurements of \citet{lyman14} are mostly on the same spectra as ours, we can simply assume that their velocity measurements have the same uncertainty; in that case their agreement with our values is even stronger, such that all their measurements agree with our values within each measurement's uncertainty.

\subsection{Different methods yield different velocities}\label{specline_94Isubsec}

\begin{figure}[!ht]    
\vspace{-.5in}
\hspace{-.8in}
\includegraphics[scale=0.42,angle=0]{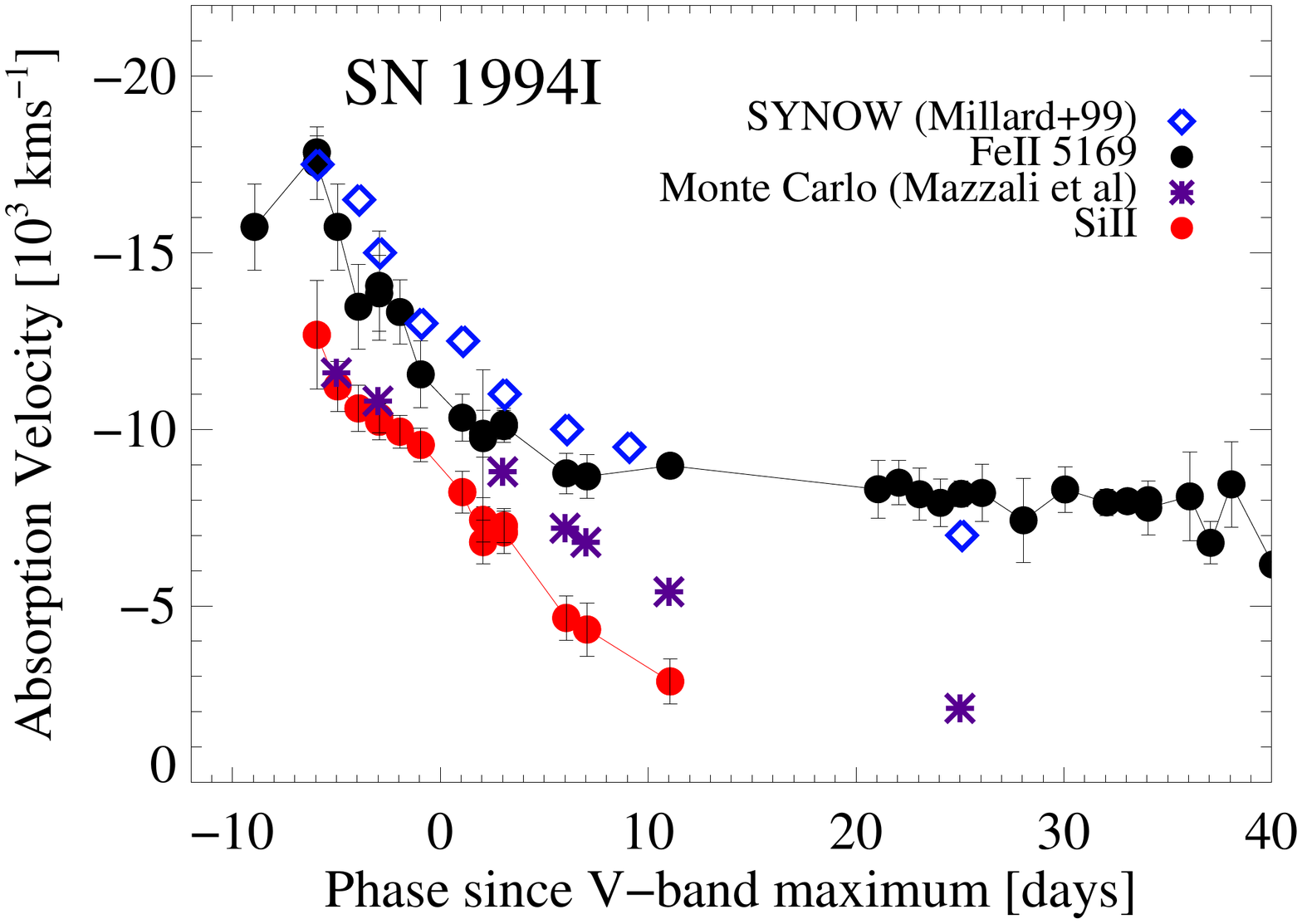}
\vspace{-.5in}
\caption{Velocity evolution of different lines and using different methods for the same SN, SN~1994I, using the same data, in almost all cases. The absorption (i.e., Doppler) velocities values for \SiSix\ and \FeFive\ were measured by us, while the SYNOW photospheric velocities are taken from \citet{millard99} and the velocities as computed from the Monte Carlo SN spectrum synthesis code of \citet{mazzali08} (and reference therein) are taken from \citet{sauer06} and \citet{pian06}. Note that different methods give different values, with SYNOW-based and \FeFive\ velocities systematically higher than Monte-Carlo-based and \SiSix\ velocities. We note that there is some controversy whether the feature at 6100$-$6200 \AA\ is purely due to \SiSix\ \citep{parrent16}.  } 
\label{sn94Ivels_fig}
\end{figure}    

Since it is not widely known that different absorption lines yield different Doppler velocities because of optical depth and ionization effects (e.g., \citealt{mazzali13}), here we use SN~1994I to show the range and types of differences, since this SN was extensively analyzed by different groups, including theoretical modeling by various groups (e.g., \citealt{baron96,baron99} and see below).

In Fig.~\ref{sn94Ivels_fig}, we plot two velocities computed from modeling in the literature, and absorption Doppler velocities measured by us for two different lines in SN~1994I, all based on the same or similar data:  photospheric expansion velocity as computed from the SN Monte Carlo spectrum synthesis code (\citealt{mazzali00}, \citealt{mazzali08} and references therein) as applied to SN 1994I (as presented in \citealt{sauer06,pian06}), the SYNOW photospheric spectra from \citet{millard99}, and the absorption velocities for \FeFive\  and \SiSix, as measured by us. In terms of the various spectral synthesis codes, different codes have different degrees of simplifications and assumptions, which is beyond the scope of this paper to describe in detail. Broadly speaking, certain codes such as the simple SYNOW code assumes a sharp photosphere, though SYNOW includes accommodating "detached" layers for certain elements (e.g. H or He) that may move at higher velocities.

The data we used for measuring the two absorption lines, namely the Berkeley (\citealt{filippenko95}) and \citetalias{modjaz14} spectra, constitute all or almost all of the same data as used for modeling in \citet{millard99} and in \citet{sauer06}. The highest velocities come from the SYNOW models, as well as from the absorptions velocities of \FeFive\ , consistent with the suggestion of \citet{branch02} that, based on SYNOW, the \FeFive\ line is a good tracer of the photospheric velocity of the ejecta. The MC photospheric velocities are lower than the aforementioned two - and importantly, they have a different velocity evolution shape, as they do not plateau at late times like the SYNOW and \FeFive\ velocities, but continue to decrease rapidly. Thus, at later times ($\Delta~ t_{Vmax}>$ +26 days) the difference is up to 7000 \kms between MC velocities and both SYNOW and \FeFive\ velocities. The \SiSix\ velocity behavior is similar to that of the MC velocities, but has a slight offset to lower values (by $\sim$2000 \kms). Thus, velocities based on SYNOW code and \FeFive\ measurements have a similar evolution on the one hand, while on the other hand, those based on the MC code and \SiSix\ measurements have another evolution. This explains why \citet{tanaka09_08Dearly}, for example, claim that for SN~2008D, the \SiSix\ line (and not the  \FeFive\ line) is tracing reasonably closely the position of the photosphere, since they compute photospheric velocities using the MC code, not SYNOW. 

The  offset between SN~1994I's MC photospheric velocities and measured \SiSix\ velocities (such that the MC photospheric velocities are higher by $\sim$2000 \kms\ than the \SiSix\ velocities) was already noted in \citet{sauer06} (their Fig. 5). They suggested that the velocity offset could be due to two reasons: the line at $\sim$6100 \AA\ (assumed be solely due to \SiSix ) may have significant contribution from Ne I 6402, such that the measurements do not trace \SiSix\ well; or there is a real discrepancy between measured \SiSix\ velocities and the computed photospheric velocities. 

While we performed this analysis explicitly only for SN~1994I, which indeed may not be a typical SN Ic (see above), the same findings hold when comparing the photospheric velocities computed by the MC code of \citet{mazzali08} (for a range of SN-GRBs, as recently presented in \citealt{walker14} and references therein) and the measured Fe velocities for the same SN (e.g., \citealt{schulze14} and as presented here in Section~\ref{Fevabs_subsec}): the MC-based velocities are systematically lower than the measured Fe absorption velocities. 

While we are not trying to resolve the issue of which line to use, nor advocate for using a specific method, we do recommend that any meaningful comparison of velocities between different SNe and different kinds of SNe Ic should be done using the same method and the same line for velocity determination. So far, modeling has been done mostly for the SN-GRBs, while \SiSix\ or \FeFive\ measurements have been reported for SN Ic in the literature, and only \SiSix\ measurements for SN-GRBs since all other lines appear to be too blended. Especially \SiSix\ vs. \FeFive\ velocities should not be used interchangeably, given their large offset from each other (up to 7,000 \kms\ at 2$-$3 weeks after maximum) and neither should the "photospheric" velocities derived by SYNOW vs. the MC code, given the systematic offset amongst the two.


\subsection{Absorption velocities of SNe Ic, SNe Ic-bl with and without GRBs}\label{Fevabs_subsec}

We concentrate on measuring \FeFive\ velocities in the same consistent way for all members of the SN Ic family: regular SNe Ic, broad-lined SNe Ic without GRBs, and SN-GRBs. The goal is to quantify in a consistent way by how much the absorption velocities of SNe Ic-bl are higher than those of SNe Ic, and if there are systematic differences for SNe Ic-bl with and without GRBs. Spectral synthesis studies suggest that Fe lines are good tracers of the photospheric velocity, since they do not saturate \citep{branch02}.

\begin{figure*}[tb!]
\begin{center}
\vspace{-.5in}
\hspace{-.3in}
\centerline{
\includegraphics[width=1.4\columnwidth,angle=0]{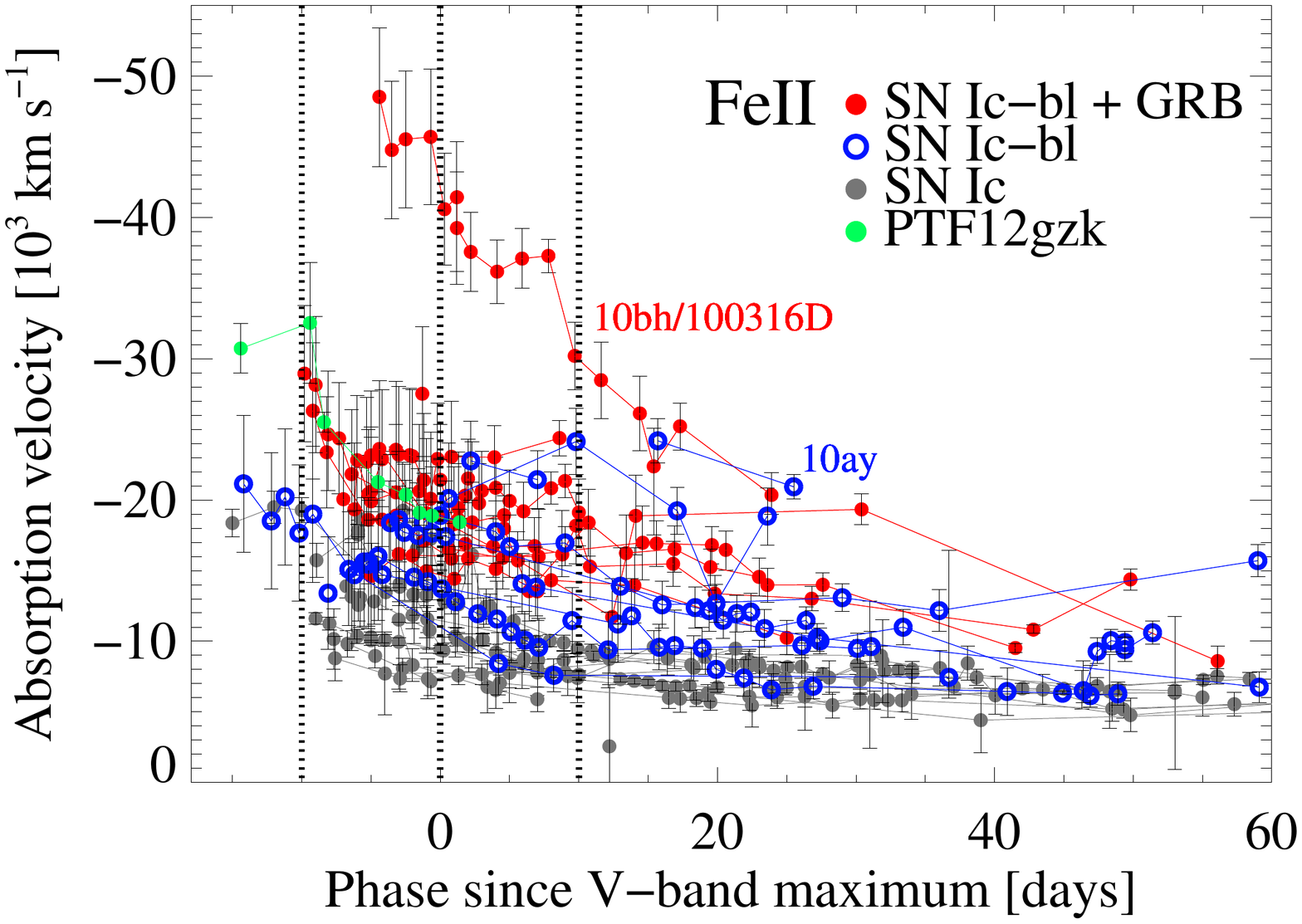}
\hspace{-1.in}
\includegraphics[width=1.4\columnwidth,angle=0]{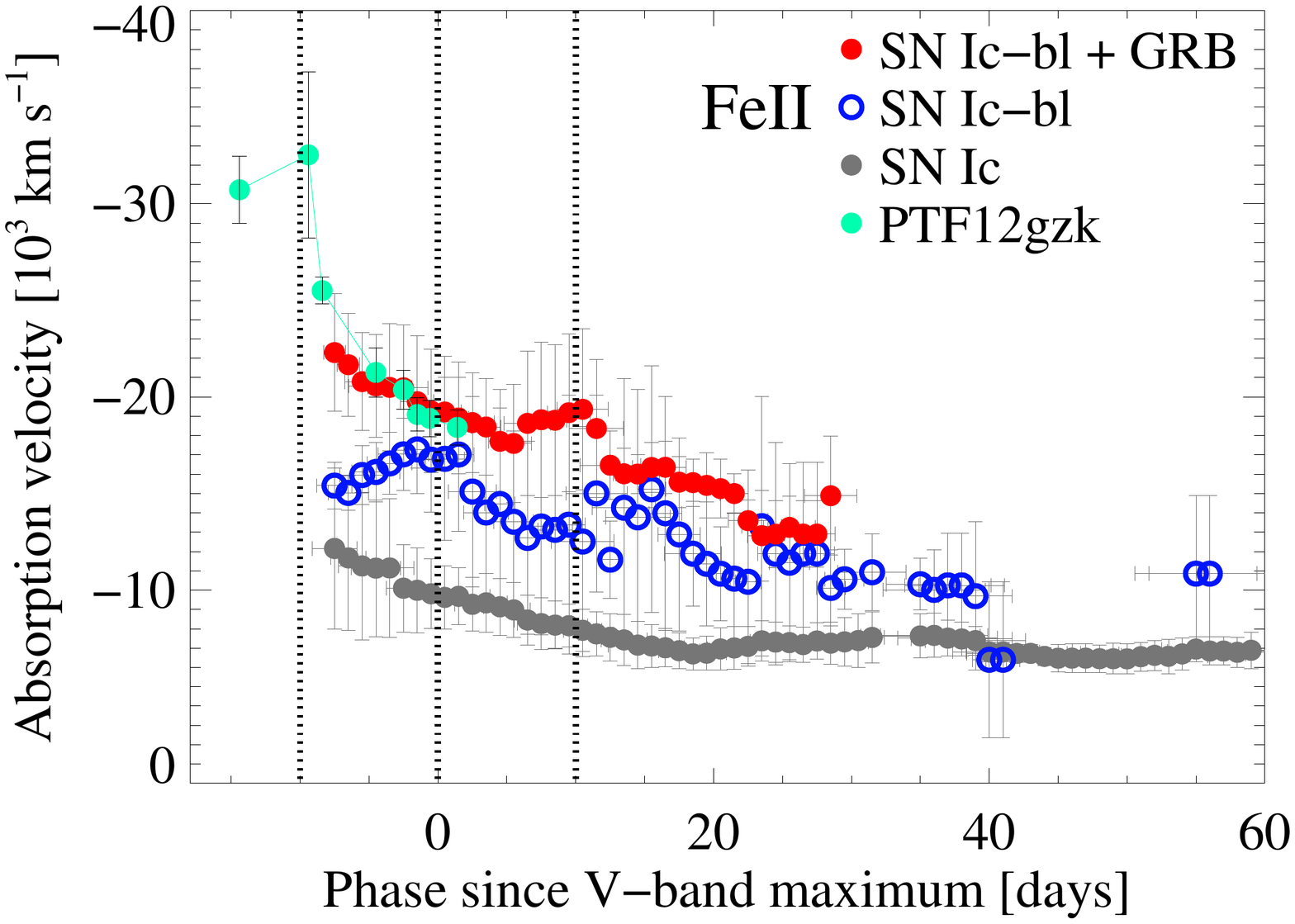}
}
\vspace{-.5in}
\caption{ \FeFive\ absorption velocities of different kinds of SNe, color-coded by SN type: SNe Ic-bl with observed GRBs, including the radio-relativistic SN2009bb, are denoted by filled red circles, SNe Ic-bl (without observed GRBs) by open blue circles, SNe Ic by filled grey circles. In addition we also show the velocity evolution of PTF 12gzk, which does not exhibit broad-lined spectra, but nevertheless shows very high absorption velocities, comparable to those of SNe Ic-bl. \emph{Left}: measured values for individual SNe.  \emph{Right}: the rolling weighted mean of the individual SN measurements for their SN subtype, not including \snbh\ nor PTF~12gzk, in rolling window sizes of 5 days (for phases with $t_{Vmax} < 30$ days) or of 10 days (for phases with $t_{Vmax} > 30$ days), in steps of 1 day. The gaps are due to the fact that no weighted average was computed for fewer than three data points in the specific windows. The error bars for the weighted mean values are the weighted standard deviations of the data. The x-axis error bars on the phases indicate the average of the phases for the velocity measurement averaged in that windows. For reference, the phases $+$10, 0 and $-$10 days with respect to $V$-band maximum are marked. SN Ic-bl with GRBs have systematically higher velocities, followed by SN Ic-bl without observed GRBs, which have lower velocities, while SN Ic have systematically the lowest velocities - this conclusion holds even when excluding SN2010bh/GRB100316D (see right plot), which appears to exhibit the highest velocities by more than 150\% than any other SN Ic-bl with a GRB. SN~Ic-bl~2010ay is discussed in the text.}
\label{Vel_comp_fig}
\end{center}
\end{figure*}

\begin{deluxetable*}{lll}
\tablecolumns{3}
\tabletypesize{\scriptsize}
\tablecaption{Weighted mean absorption velocities  }
\tablehead{
\colhead{SN Type } &
\colhead{$v_{abs}$(\FeFive)\tablenotemark{a}  [\kms ]  } &
\colhead{$v_{abs}$(\FeFive) [\kms ] } \\
\colhead{} & 
\colhead{at $t_{Vmax}$=0 days} & 
\colhead{over all epochs  } 
}
\startdata
SN Ic\tablenotemark{b}        &    $-$9,700 $\pm$ 1,500   &  $-$8,900 $\pm$ 3,300  \\  
SN Ic-bl (all)              &    $-$21,100 $\pm$ 8,200  &  $-$18,000 $\pm$ 8,000 \\
&  \\
SN Ic-bl (no GRBs\tablenotemark{c})    &    $-$16,700 $\pm$ 2,700  & $-$13,300 $\pm$ 4,600 \\ 
SN Ic-bl + GRBs        &    $-$23,800 $\pm$ 9,500 & $-$21,000 $\pm$ 8,000 \\
\hline
  & \\
\multicolumn{3}{c}{When excluding \snbh\  } \\
SN Ic-bl (all)              &    $-$18,500 $\pm$ 3,300  &  $-$16,200 $\pm$ 5,200   \\
SN Ic-bl + GRBs     &     $-$19,200 $\pm$ 2,900  & $-$17,000 $\pm$ 4,000 \\
\enddata
\tablenotetext{a}{The weighted means were taking over a 5 day rolling windows centered at $t_{V}$= 0, i.e. $V$-band maximum. The error bars refer to the weighted standard deviation of the data. For multiple velocity measurements for the same SN across the 5-day window, the weighted average for that SN was taken, as to not bias the mean SN type values by a few SNe that have many data points. Note that PTF~12gzk is not included in the mean values for SNe Ic.} 
\tablenotetext{b}{Excluding PTF12gzk.}
\tablenotetext{c}{While no GRB emission was observed for these SNe Ic-bl, the necessary data to rule out an off-axis and/or low-luminosity GRBs are present for some, but not for all SNe Ic-bl in this sample.}
\label{wmeanvels_table}
\end{deluxetable*}

Figure~\ref{Vel_comp_fig} shows the \FeFive\ absorption velocities of \nSNGRB\ SN-GRBs, \nSNIcbl\ SNe Ic-bl and \nSNIc\ SNe Ic, color-coded by SN subtype. This is the most detailed and consistent analysis on the largest data set to date. The left panel displays the measurements for the individual SNe, while the right panel displays the rolling weighted mean velocity evolution for each SN type in order to display their bulk evolution. We follow our procedure from \citet{liu15} to construct rolling weighted means, in order to mitigate the impact of the choice of the beginning and end points of the bin by replacing the bin with a rolling window. To summarize, we took the weighted average within a certain window size and moved, or "rolled",  in 1-days steps. We chose the rolling window size for early times (before 30 days after max) to be 5 days, as the velocity evolution is fast at early times, and for phases after that, we chose a 10-day-window. In order to mitigate the impact of a single SN that has many data points over the window range, we took the weighted average of the data points for the same SN (within the window range) first. We then computed the weighted mean absorption velocity of all SNe of the same type within that rolling window if there were more than three SNe with data in that window (for any standard deviation to be meaningful). For the center phases of the rolling window, we took the mean of the phases at which the velocity measurements were made that contributed to the weighted mean velocity. Obviously, the rolling weighted mean velocities are not independent of each other, as adjacent windows may include many of the same SN data.

The following was not encountered in \citet{liu15} but is addressed here: for the asymmetric errors of Doppler absorption velocities for the SNe Ic-bl (derived from the template fitting method via MCMC which outputs the full posterior probability density function), we followed the prescription of \citet{barlow03} to construct weighted averages from asymmetric uncertainties (their model 1, though there was no significant difference between using their model 1 vs. their model 2). The error bars for the weighted mean values are the unbiased weighted standard deviation of the data (not that of the weighted mean), since we would like the error bar to reflect the variation of the velocity evolution in the SNe, not just the quality of their velocity measurements (i.e., the uncertainty in the velocity measurements).  In Table~\ref{wmeanvels_table}, we list the weighted mean velocities for both lines at $V$-band maximum, in a window of five days.

Now, for the first time, we have a large enough sample analyzed in a systematic manner to quantify the similarity and behavior of SNe Ic-bl (especially those accompanied by GRBs) compared to those of SNe Ic - something that has been attempted before but only by using SN~1994I as a representative of all SNe Ic (e.g., \citealt{pian06,valenti08,taubenberger06}). A number of important trends can be seen in Fig.~\ref{Vel_comp_fig}, where in the following the quoted weighted mean velocities are computed for $V$-band maximum in a five-day window, and where the error bar is the standard deviation in the data: \\

$\bullet$ When grouping together all SNe Ic-bl (both with and without GRBs) in one category, the weighted mean velocity is  $<v_{SNIc-bl}>= -22,200~\pm$ 9,400 \kms\ . This is more than a factor of two above the value for SNe~Ic (weighted $<v>_{SNIc}= -9,700 \pm$ 1,600)\footnote{We note that our conclusions hold even if our new method of estimating absorption velocities for SN Ic-bl via MCMC template fitting (Appendix~\ref{speclineID_sec}) is not correct, since the velocities measured via the classical identification methods are even higher than those from template fitting.}. We note that the absorption velocities of \snbh\ are the highest overall for any SN-GRB, higher by $\sim$ 150\% compared to any other SN Ic-bl with a GRB at $V$-band maximum. This was already noted by \citet{schulze14}, however, their measured values of the other SN-GRBs velocities were also high, since they were using the more simplistic identification method (Section~\ref{litcomparison_subsec}), which tends to overestimate absorption velocities for when \FeFive\ is highly blended, as in SNe Ic-bl. In case \snbh\ is different from the rest of the SNe Ic-bl, we exclude it when we list the weighted mean velocity of all SNe Ic-bl in Table~\ref{wmeanvels_table}. We find that the weighted mean velocity for all SNe Ic-bl is still almost a factor of two higher than the weighted mean velocity for normal SNe Ic.

$\bullet$ We find that SNe Ic-bl \textbf{with} GRBs have, on average, \emph{consistently} \textbf{higher} \FeFive\ absorption velocities than SNe Ic-bl \textbf{without} GRBs (by $\sim$3,000 \kms\ for the date of $V$-band maximum, and by  $\sim$6,000 \kms as an average over all phase, for both cases when excluding \snbh , see Table~\ref{wmeanvels_table}).  SNe Ic have the lowest absorption velocities.  Nevertheless, as the plots indicate, there is an overlap in absorption velocities between the different types of SNe, though the systematic trend of highest mean velocities for SN-GRBs and lowest mean velocities for SNe Ic is present at almost all epochs.

$\bullet$ The biggest difference in mean velocities for the different sub-types of SNe Ic is at the earliest phases (at $t_{Vmax}= -10$ days) and the difference decreases with time. Indeed, after $t_{Vmax}$=30 days the Doppler absorption velocities of all three subgroups become similar and are consistent within the standard deviation of the data. In general, SNe Ic have the smallest dispersion (excluding PTF~12gzk), especially at later phases, while SN-GRBs have the largest dispersion - however since that dispersion is mostly driven by  \snbh\ , when discounting that SN-GRB the group with the largest dispersion (driven by the behavior of many SNe) is the group of SNe Ic-bl without observed GRBs. This is the case at many phases, especially at 0$ < t_{Vmax} < $ 30 days.

$\bullet$SN Ic-bl 2010ay shows high \FeFive\ absorption velocities, much higher than those for SNe Ic-bl without observed GRBs (see also \citealt{sanders12_10ay}).  While no gamma-rays were detected from SN~2010ay, the gamma-ray and radio upper limits cannot rule out the possibility that it hosted a common and low-luminosity GRB like GRB060218 \citep{sanders12_10ay}. In fact, many of the properties of SN~2010ay are similar to those of SN-GRBs: its high absorption velocity (shown here), its high peak luminosity, and its low metallicity \citep{sanders12_10ay}. Thus, we speculate that SN~Ic-bl 2010ay is a very good candidate for a low-luminosity GRB, similar to \snaj , seen either on- or off-axis. \\

 \subsection{Line Widths in SNe Ic-bl with and without GRBs}\label{Fewidths_sec}

\begin{figure}[!ht]    
\vspace{-1.8in}
\hspace{-.7in}
\includegraphics[scale=0.54,angle=0]{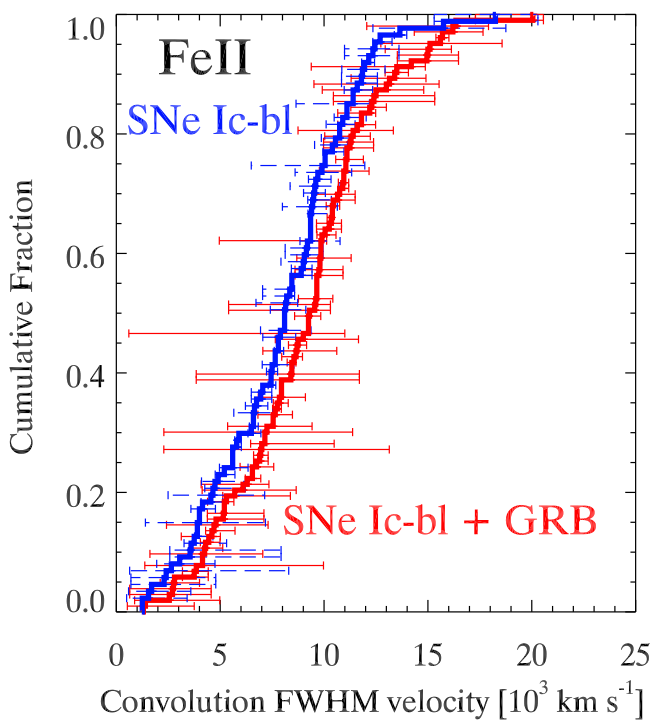}
\caption{Cumulative fraction (solid lines) of  \FeFive\ convolution FWHMs (based on the $sigma$ velocities, see Appendix~\ref{speclineID_sec}), with respect to the appropriate SN Ic template, of SN Ic-bl with observed GRBs (in red) and without observed GRBs (in blue), and their error bars on the velocities. These convolution FWHM velocities with which the template SN Ic had to be convolved to match the individual SNe Ic-bl spectra indicate the width of the spectral line \emph{with respect} to the SN Ic template. SN Ic-bl with GRBs have higher $\sigma$ velocities, i.e. larger \FeFive\ line widths than SN Ic-bl without GRBs. A K-S test yields the small probability that 0.8\% that the FWHM convolution velocities of SN Ic-bl without GRBs are drawn from the same parent population as SN Ic-bl with GRBs. }
\label{SNIcbl_sigmaCDF_fig}
\end{figure}

We present a novel approach to characterize and quantify the ``broadness" of lines in SN Ic-bl spectra. Besides the Doppler velocity, $v$, our template fitting procedure for SNe Ic-bl (Appendix~\ref{speclineID_sec}) also outputs the width of the Gaussian,$\sigma$, with which the appropriate SN Ic template has to be convolved to match the observed SN Ic-bl spectrum. We converted $\sigma$ to FHWM (which is straightforward for a Gaussian) and call it the ``FWHM convolution" velocity. It thus indicates the width of the SN Ic-bl spectral line \emph{with respect} to the SN Ic template at the $\sim$same phase.
While the absolute width decreases as a function of time for the same SN (similar to the absorption velocities decreasing in time), we find that there is no decrease in the convolution FWHM as a function of time, since these values are with respect to a SN Ic template, whose width decreases with time. Thus, since there is no time evolution, we plot all convolution FWHM values of SNe Ic-bl with and without GRBs in Fig.~\ref{SNIcbl_sigmaCDF_fig} in terms of cumulative fractions, and also plot the (asymmetric) error bars as horizontal lines. We report median values in Table~\ref{convolutionvels_table}.

\begin{deluxetable}{ll}
\tablecolumns{2}
\tabletypesize{\scriptsize}
\tablecaption{Median convolution FWHM velocities}
\tablehead{
\colhead{SN Type } &
\colhead{$v_{\mathrm{convolution FWHM}}$(\FeFive)\tablenotemark{a} } \\
\colhead{} & 
\colhead{ [\kms ]} 
}
\startdata
\multicolumn{2}{c}{Median values based on all data points\tablenotemark{b}} \\
SN Ic-bl (no GRBs\tablenotemark{c})     &   8,200 (N=87)   \\
SN Ic-bl + GRBs        &   9,300  (N=103) \\
\hline
 & \\
\multicolumn{2}{c}{Median values based on the weighted mean for each SN } \\
SN Ic-bl (no GRBs\tablenotemark{c})     & 7,600 (N=10)   \\
SN Ic-bl + GRBs        &  9,300 (N=9)   
\enddata
\tablenotetext{a}{These convolution FWHM velocities are with respect to the the width of the SN Ic template at the appropriate phase.} 
\tablenotetext{b}{May include multiple data points per SN.}
\tablenotetext{c}{While no GRB emission was observed for these SNe Ic-bl, the necessary data to rule out an off-axis and/or low-luminosity GRBs are present for some, but not for all SNe Ic-bl in this sample.}
\label{convolutionvels_table}
\end{deluxetable}

We find that SNe Ic-bl (both types) have \FeFive\ widths that are around 8,800 \kms\ (median value) broader than those of SNe Ic (based on data from all epochs). In addition, figure~\ref{SNIcbl_sigmaCDF_fig} shows that SN-GRBs have consistently higher convolution \FeFive\ FWHM velocities than SNe Ic-bl without GRBs by 1,000 $-$ 2,000 \kms (see Table~\ref{convolutionvels_table}) - the same trend as for absorption velocities (Section~\ref{Fevabs_subsec}). A Kolmogorov-Smirnov (KS) test indicates that there is a small chance, namely 1.8\% that the FWHM convolution velocities of SN Ic-bl without GRBs are drawn from the same parent population as SN Ic-bl with GRBs. As before there may be the concern that for this approach, namely including all available data points for each for each SN, would bias the results towards the SNe with many data points. Thus, using the same procedure as in Section~\ref{Fevabs_subsec}, we took the weighted average of all data points for each SN and then computed the median for all SNe of the same type. The results are unchanged, and the result of the KS test is 0.3\%. 
To get an estimate on how the velocity error bars (shown in Fig.~\ref{SNIcbl_sigmaCDF_fig}) may influence the results, we compute the KS probability for 2 extreme scenarios using data at all phases: when the two velocity distribution are closest  (i.e., adding the errors to values of the SN Ic-bl without GRBs and subtracting the errors from the SN-GRB values), and when the two are furthest. For the scenario where they are closest, the KS test still obtains a low probability of 3\%. Thus, even including error bars and the most extreme case, the line widths of SN-GRBs and those of SN Ic-bl without observed GRBs are unlikely to be drawn from the same parent population. Of course, for the scenario when they are furthest, the KS test obtains the 0\% probability within our significance. 

In summary, we conclude that the \FeFive\ widths of SN-GRBs are higher than those of SNe Ic-bl without observed GRBs (by $\sim$ 1,000$-$2,000 \kms), in line with the visual trend for all features in the mean spectra of SNe Ic-bl with and without GRBs (Section~\ref{average_subsec}, Fig.~\ref{meanspec_SNIcbl_withwithoutGRB}) and similar to the trend seen for \FeFive\ absorption velocities (Section~\ref{Fevabs_subsec}). We discuss the implications of this finding, together with our speculations, in the next section.

  \subsection{Absorption Velocities vs. Line Width Velocities}\label{Fewidths_sec}

\begin{figure}[!ht]    
\vspace{-.5in}
\hspace{-.8in}
\includegraphics[scale=0.42,angle=0]{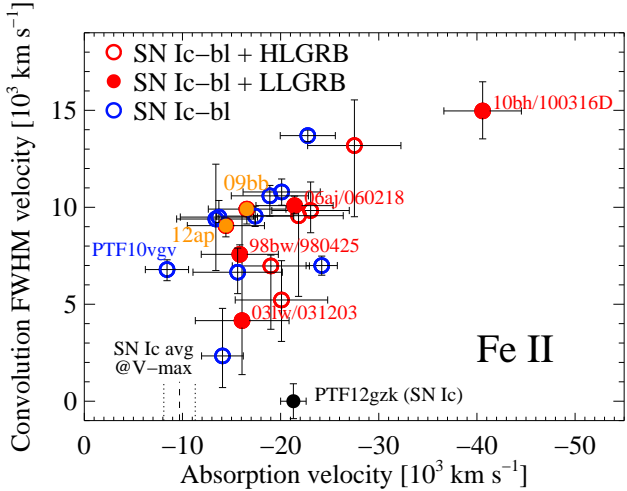}
\vspace{-.5in}
\caption{Comparison between \FeFive\ convolution FWHMs and absorption velocities with respect to the appropriate SN Ic template, of SNe Ic-bl with observed GRBs (in red; in addition, filled circles: with LLGRBs and their names reported; open circles: with HLGRBs) and without observed GRBs (in blue) for the spectrum that is closest to the respective SN date of $V$-band maximum. These convolution FWHM velocities with which the template SN Ic had to be convolved to match the individual SNe Ic-bl spectra indicate the width of the spectral line \emph{with respect} to the SN Ic template. For reference we also mark the weighted average absorption velocity for SNe Ic (dashed line), as well as its standard deviation (dotted lines), and its convolution velocity of 0 $\pm$ 1000 \kms. While absorption velocity and width velocity are correlated for many SNe, there are also outliers, such as PTF~12gzk, which has narrow lines (it is a SN Ic), but high absorption velocity, and PTF~10vgv, whose absorption velocity is lower than the average for SNe Ic, but still is a broad-lined SN Ic. We also plot in orange the values for the radio-relativistic SNe 2009bb and 2012ap. See text for discussion on any differences between LLGRBs and HLGRBs.}
\label{vabs_vfwhm_fig}
\end{figure}

For all SNe Ic-bl in our sample, we plot the absorption velocity against the line-width velocity based on the same spectrum, measured as close as possible to $t_{Vmax}$ (Fig.~\ref{vabs_vfwhm_fig}). We note that for some SNe, the spectrum closest to $t_{Vmax}$ is actually late, e.g., at $t_{Vmax}$=15.7 days for SN~2010ay. While in many cases, high absorption velocities correlate with broad lines (the Spearman coefficient is $\rho=-0.599$ with a strong significance of 0.004, and the Pearson coefficient is $r = -0.642$), there are two SNe for which that is not the case (SNe PTF10vgv and PTF12gzk). This is the first time such a correlation has been investigated explicitly and methodically for a large sample of SNe Ic-bl.

As shown, PTF~12gzk possesses very high absorption velocities,\footnote{Radio observations suggest that the synchrotron radiation emitting layer, i.e. the fastest-moving ejecta, also possessed high velocities, with a mean velocity of 80,000 \kms \citep{horesh13}.} well within those of SN Ic-bl with GRBs, but narrow lines (similar to those of SNe Ic, see also \citealt{benami12}). In terms of reasons for the mismatch between the two kinds of velocities for PTF~12gzk,  \citet{benami12} speculated that it could be due to aspherical explosion geometry, or high ejecta mass or very steep density gradient. PTF10vgv, on the other hand, possesses broad lines, but has the lowest absorption velocity amongst all SNe Ic-bl, lower by a factor of  $\sim$1.5 than the average value for SNe Ic-bl, and even lower than the average for SNe Ic. This must be the reason why \citet{corsi11} classified this SN as a SN Ic, since their definition of SNe Ic-bl relies on the value of the ``photospheric" velocity (in their case, inferred from spectral synthesis modeling based on the code from \citealt{mazzali00}), which is low for PTF~10vgv. However, the definition of a broad-lined SN Ic relies on the measured width of the line and should be adopted for classifying SNe.

We highly encourage modeling the spectra of SNe Ic-bl with a detailed parameter study in terms of e.g., asphericities, density gradients and ejecta masses, to reproduce the observed parameter space of all SNe Ic-bl and not to just fit a hand-crafted model spectrum to an individual SN.

In addition, we also search for differences in SN Ic-bl spectra for those connected with Low-Luminosity GRBs vs those connected with High-Luminosity GRBs, as some theories suggest different physical mechanisms for LLGRBs and HLGRBs. Figures \ref{vabs_vfwhm_fig} and \ref{fig_absvels_sntypes} suggest no obvious trend between the SN spectral property (specifically, $v_{abs}$) and GRB luminosity. While SNe Ic-bl connected with LLGRBs appear to show a larger range in velocities ($-16,000 < v_{abs} < -40,0000$ \kms\ at $V$-band maximum) than SN Ic-bl connected with HLGRBs ($-19,000 < v_{abs} < -27,0000$ \kms\ at $V$-band maximum), this is only due to one object, \snbh, which has velocities that are much larger, up to 150\%, compared to the rest of SN-GRB. In other words, for velocities at $V$-band maximum, the standard deviation of the HLGRB sample is $\sim$ 3300 \kms, while that of the LLGRB with \snbh\ is $\sim$10,000 \kms, however without it, the standard deviation drops to $\sim$2500 \kms, so becomes similar to that of the HLGRB sample. Given the small data set of a total of four objects, it is hard to determine whether \snbh\ is an outlier, and thus more data are needed. 

We further discuss the implications of this work in Section~\ref{discussion_sec}, and specifically in \ref{llgrbvshlgrb}.

\section{Discussion: Constraints on Progenitors and Explosion mechanisms}\label{discussion_sec}
Although only SNe Ic-bl have been connected with GRBs, the majority of SNe Ic-bl show no detectable gamma-ray emission, and their relationship to normal SN Ic is unknown. Thus, one of the key outstanding questions is what conditions lead to each kind of explosion in which kinds of massive stripped stars. We have explored this question by investigating what sets apart the spectra of SNe Ic-bl with GRBs from those without GRBs, and how SNe Ic-bl spectra are related to those of SNe Ic. We have uncovered a number of trends while analyzing our datasets. Here, we discuss those that are specifically related to constraining progenitors and explosion mechanisms.

\subsection{Helium in Progenitors of SN Ic-bl and of SN-GRBs?}
In Section~\ref{he_subsec}, we addressed the question whether He lines could be smeared out in the spectra of SNe Ic-bl such that they would not be detectable, despite He being present in the ejecta. The lack of He-detection in optical spectra of SNe Ic-bl and GRB-SNe is puzzling, since many of the progenitor models, especially those of SN-GRBs, are supposed to be He-stars. Here, we simulated a ``broad-lined SN Ib", i.e., a broadened SN Ib with He lines. While convolving a SN Ic mean spectrum with a Gaussian profile yields a spectrum that is very similar to a SN Ic-bl mean spectrum, a similarly convolved SN Ib mean spectrum (effectively a ``SN Ib-bl") is different from the SN Ic-bl spectrum by more than 2$\sigma$ over the wavelength ranges of interest. We tested this for the most extreme cases of smearing (i.e., for large convolution velocities). Thus, we conclude that the absence of clear He lines in optical spectra of SNe Ic-bl and of SN-GRBs is not due to the smearing out by the high velocities present in SNe Ic-bl. This implies that the progenitor stars of SN-GRBs may be truly free of He, in addition to being H-free, putting strong constraints on the stellar evolutionary paths needed to produce such GRB and SN Ic-bl progenitor stars at the observed low metallicities (e.g, \citealt{modjaz08_Z,sanders12,graham13}), including the potential need for enhanced convection to bring He into deeper layers and therefore burn it \citep{frey13}. 

Since standard mechanisms in stellar evolutionary models for single stars find it very difficult to remove so much He such that no or very little He (less than 0.2 \msol\ ) remains (e.g.,  \citealt{yoon10,yoon15}), especially at the lower metallicities, at which SN-GRBs are generally found, one of the only remaining paths may be rapid mass loss after core helium exhaustion due to interaction with a compact companion in a very short-period binary system (S-C. Yoon, private communication, \citealt{pols02,ivanova03}) in potentially young dense star cluster \citep{vandenHeuvel13} or very massive He-stars (\citealt{yoon10}, see their Fig. 10, but only for solar metallicity). Another viable alternative is explosive common-envelope ejection in a slow merger of a massive primary with a low mass companion \citep{podsiadlowski10}. Thus, our He-constraint supports short-period binary models for SN-GRBs. This suggestion is in line with recent observational work on host galaxies of SNe Ic-bl, SN-GRBs and GRBs by \citet{kelly14}, who interpret their preference for over-dense galaxies as indicating that they preferentially come from tight binary systems, which may be produced more readily in such over-dense conditions. There is yet another mechanism to ``hide" He in SN Ic spectra: if the amount of $^{56}$Ni mixing is not sufficient, then the He lines will not be visible in the optical spectra \citep{lucy99,hachinger12,dessart12}. However, based on our thorough analysis of the largest data set of SNe Ib and SNe Ic in \citet{liu15}, we think that there is no hidden He in normal SNe Ic , since the SN observables are inconsistent with the predictions of \citet{dessart12}, who had modeled the effect of insufficient mixing in SNe Ic spectra that may have lead to hiding He.  Similar conclusions were reached by \citealt{cano14_99dn} based on modeling of SN Ib 1999dn and SN Ic 2007gr. \\ 

Nevertheless, in order to fully understand the progenitors of SNe Ic-bl, we highly encourage modeling the spectra of SNe Ic-bl in detail to understand the physical conditions and to do a detailed parameter study, which has not been done before, using sophisticated spectral synthesis models such as CFMGEN \citep{dessart11}, PHOENIX \citep{baron95}, and SEDONA \citep{kasen06}. In this context, we note that there are discrepancies between the simulated SN Ic-bl spectrum (broadened SN Ic mean spectrum) and the real SN Ic-bl spectrum for two different regions: a) in the blue region (3500-4400 \AA, Fig.~\ref{meanspec_SNIcbl_withwithoutGRB}), where the forest of Fe lines lie, which may have implications for understanding line formation in highly-expanding atmospheres; and b) the region around 6300\AA\ (with absorption commonly associated with \SiSix ). In addition, we find that high-absorption-velocity SNe may not have broad lines (e.g., PTF12gzk), and conversely, that broad-lined SNe Ic can have low absorption velocities (e.g., PTF10vgv). Given that these two SN mavericks were found by PTF, we are looking forward to more interesting SNe discovered via untargetted and innovative surveys, such as LSST. In the meantime, until a unifying picture of spectral line formation in SNe Ic and SNe Ic-bl is developed and tested, we suggest that authors, when reporting velocities, are clear about how the number was derived (whether from modeling or observations, and which line was used) and whether it is the absorption velocity or the width of the line they are reporting, and to use the same method when comparing different SNe.

\subsection{Progenitors and Explosion Conditions of SNe Ic-bl with GRBs vs. SNe Ic-bl without observed GRBs}

We have found that SNe Ic-bl with observed jets have quantifiably broader lines (Sections~\ref{snlc_blwithandwithoutGRBs_subsec} and~\ref{Fewidths_sec}) and higher absorption velocities (Section~\ref{Fevabs_subsec}), and therefore higher ejecta velocities than those without observed GRBs. One explanation is that the latter either had lower-energy or shorter-duration jets that were stifled and could not break out of the star, but nevertheless imparted some of their energy to the stellar envelope and therefore accelerated it \citep{lazzati12}, though to lower velocities than for SN-GRBs. In this scenario, radio-relativistic SNe such as SNe~2009bb and 2012ap \citep{soderberg10_09bb,margutti14_12ap}, which are claimed to be engine-driven, but with no GRB detection, are explosions where the jet may have lasted for a time that is shorter than for the SN-GRB \citep{margutti14_12ap}, but longer than for a SN Ic-bl without a GRB. Another reason why the jet could have been choked is because the stellar envelope, which became SN ejecta when the star exploded, had a larger mass in SNe Ic-bl without GRBs, and thus gave rise to lower velocities, as observed, for the same kinetic energy as in SNe Ic-bl with GRBs. If both kinds of SNe Ic-bl had the same kinetic energy, the ejecta mass of SNe Ic-bl without observed GRBs would be on average $<v_{abs}(SNGRB)>^2/<v_{abs}(SNIc-bl)>^2  =$ 2.6 times higher compared to the ejecta mass of SN-GRBs (or twice when excluding \snbh ). This extra mass may lead to the stifling of the jet. Conversely, if both SN-GRB and SNe Ic-bl had the same ejecta masses, but released different amounts of kinetic energy, then SN-GRBs would be releasing 2.6 times more energy than SNe Ic-bl without GRBs (or twice if excluding \snbh ). At face value, spectral synthesis modeling by \citet{mazzali13} and collaborators suggests that \snbw , \sndh , \snlw\ have higher ejecta masses and higher explosion energies than SNe Ic-bl without GRBs, based on their small sample (though keeping in mind all the caveats about geometrical assumptions and simplifications). This result is in line with calculations by \citet{cano13} on a larger set of SNe Ic-bl and SN-GRBs, though there are a number of assumptions and simplifications in \citealt{cano13}, including using Fe- and Si-based velocities interchangeably, even though they give systematically different values as shown above. Thus, we conclude that it is reasonable that SNe Ic-bl without observed GRBs, which appear to have intrinsically lower kinetic-energies than SN-GRBs, may have had also lower energy jets that were stifled in the passage through the star. In addition, it is possible that both phenomena (successful jet and higher velocities in the SN ejecta) are caused by the same ingredient, such as high angular momentum (e.g. \citealt{fuller15,moesta15}). There is an assumption in some papers that the GRB jet inside the star has on average a small energy of  $\sim 10^{51}$ ergs (e.g., \citealt{mazzali14}). This is a misconception, since $10^{51}$ ergs is the average gamma-ray radiation energy of the jet after it breaks out of the star, which can be lower due to losses and inefficiencies in producing radiation. Indeed, most simulations include much stronger jets that travel through the star, almost always with energies of $10^{52}$ ergs (e.g., \citealt{lazzati13,duffell15}). 

Thus, it is reasonable to conclude that SNe Ic-bl without observed GRBs, which appear to have intrinsically lower kinetic energies than SN-GRBs, may also have had lower-energy jets that were stifled in the passage through the star, potentially more affected by the kink instability \citep{bromberg16}.

Recently, various works have suggested magnetars as SN-GRB central engines (e.g., \citealt{usov92,thompson04,bucciantini07,metzger11}), whose rotational energy reservoir of $10^{52}$ ergs (\citealt{usov92}, but see more recently \citealt{metzger15}) appears to be consistent with the high average GRB-SN energies of $\sim2 \times 10^{52}$ ergs \citep{mazzali14,cano15_14bfu}.\footnote{However, note that the KE was estimated in various different ways in \citep{mazzali14} and that the SN KE values in \citet{cano15_14bfu} are not corrected for asymmetry.} However this would imply that all SN-GRB would need to be maximally efficient in extracting the rotational energy of the rapidly rotating pulsar, which is somewhat hard to imagine given the usual inefficiencies in astrophysical systems.


\begin{figure*}[!ht]    
\begin{center}
\vspace{-.5in}
\hspace{-.3in}
\centerline{
\includegraphics[width=1.4\columnwidth,angle=0]{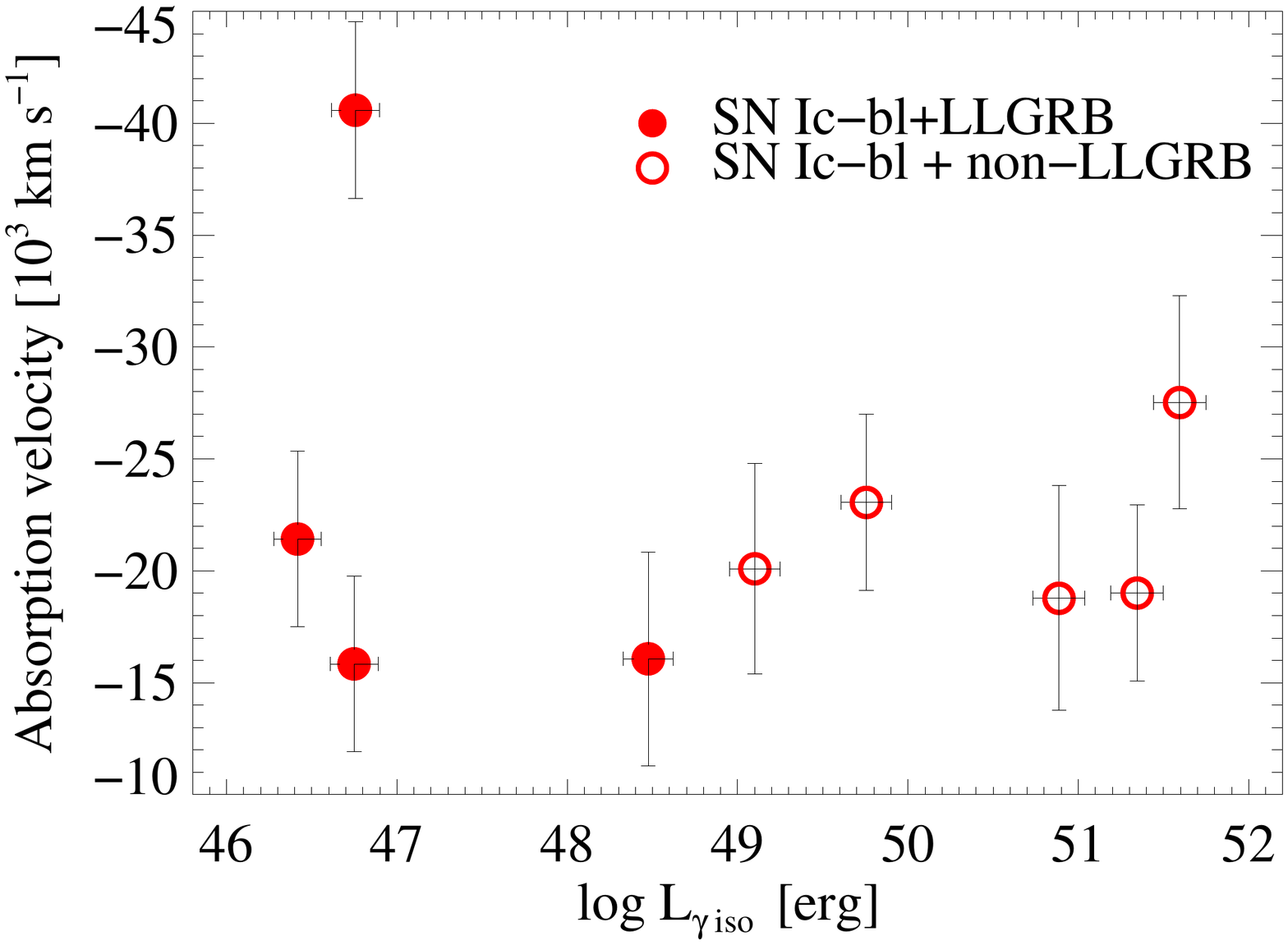}
\hspace{-1.in}
\includegraphics[width=1.4\columnwidth,angle=0]{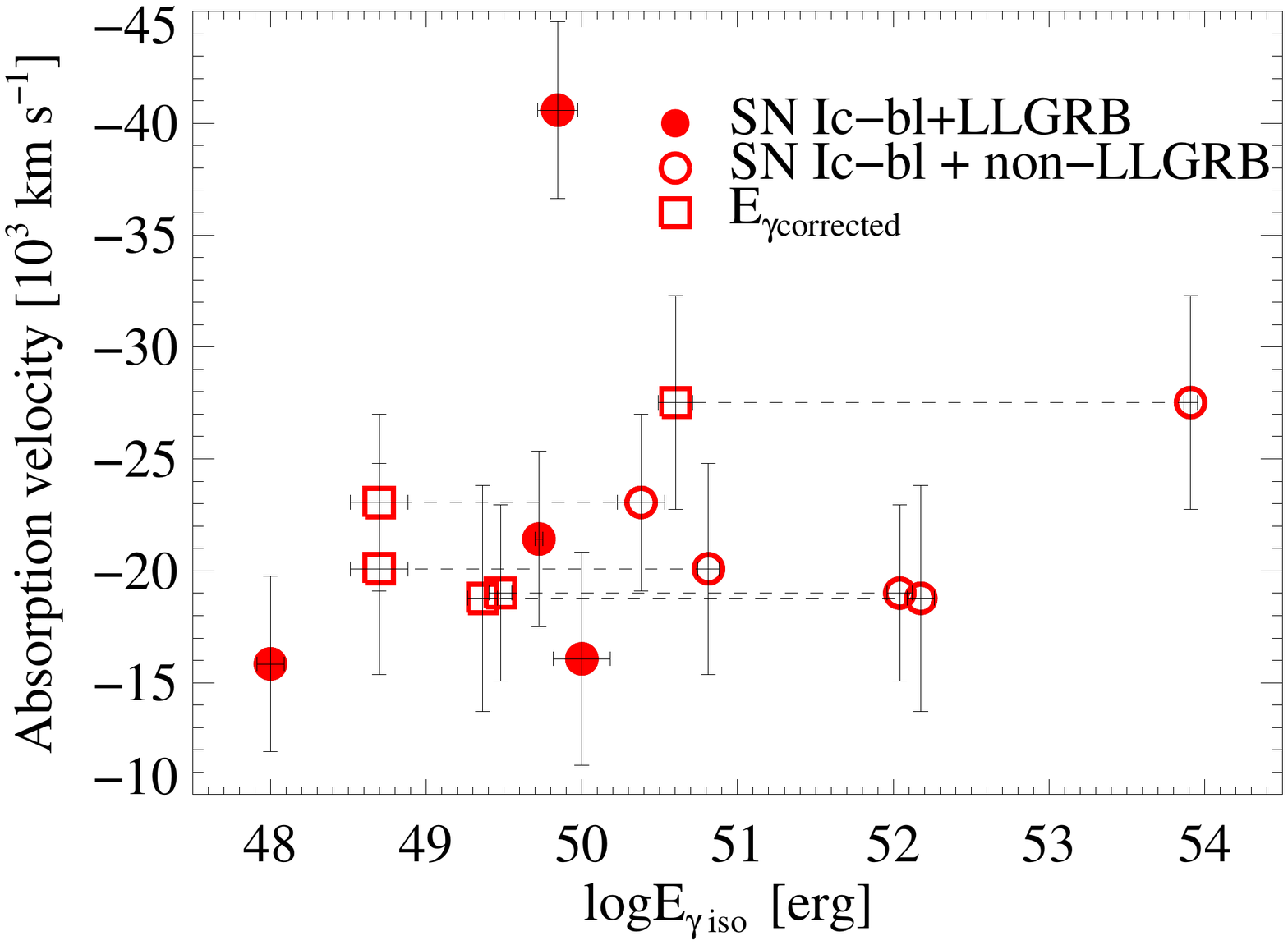}
}
\vspace{-.5in}
\caption{Comparison between SN spectral properties and the high-energy properties of their accompanying GRBs: absorption velocity (of \FeFive\ at $V$-band maximum) as a function of  isotropic gamma-ray luminosity ($left$) and of gamma-ray energy ($right$, showing both isotropic and beaming-corrected energies, connected by a dashed line for beamed GRBs). The velocities are based on the SN spectrum within $t_{Vmax}=\pm$2 days. No strong correlation between SN spectral properties and GRB properties are seen (see text for more discussion).}
\label{vabs_GRB_fig}
\end{center}
\end{figure*}    

Another potential explanation is that the lower velocities of SNe Ic-bl without observed GRBs is due to viewing angle effects, where SNe Ic-bl have no observed GRB because their beamed GRB jets were not pointing along our line of sight; such SNe harboring off-axis GRBs would presumably also have lines that are less broadened, since we would not be looking at the SN ejecta that got most accelerated by the passage of the jet along the line-of-sight. We think that it is unlikely that all nearby SNe Ic-bl without observed GRBs in our sample are due to off-axis GRBs, for the following reasons:  more than half of the nearby (z $<$ 0.3) SN-GRBs have very large opening angles of up to 180 $\deg$ (see e.g., Table 1 in \citealt{mazzali14}), such that highly collimated GRBs are probably rare in the nearby universe (e.g., \citealt{guetta07,baerwald14_grbrates}). In addition, other lines of evidence suggest that the majority of SNe Ic-bl are not off-axis GRBs: deep radio observations have not detected off-axis GRBs in any SN Ic-bl or any Stripped SN (\citealt{soderberg06_radioobs,soderberg10_09bb,corsi11,corsi12}, R. Margutti, private communication 2015) $-$ and in the cases where radio emission was detected, it was consistent with a model of the pure hydrodynamic explosion, except for the two cases of engine-driven SNe Ic-bl 2009bb and 2012ap \citep{soderberg10_09bb,margutti14_12ap}. Furthermore, SNe Ic-bl with GRBs are at systematically lower metallicities than SN Ic-bl without observed GRBs \citep{modjaz08_Z,modjaz12_proc,sanders12}, which cannot be explained by viewing angle effects.
Thus, it is unlikely that all nearby SNe Ic-bl without observed GRBs in our sample are due to off-axis GRBs, and thus we favor the first hypothesis, namely that SNe Ic-bl without observed GRBs had lower-energy or shorter-duration jets, as also supported by the computed kinetic energies of a few SNe Ic-bl, compared to those of a few SN-GRBs.

In either case, both scenarios support the hypothesis that the engines of SNe Ic-bl (with and without GRBs) are fundamentally different from those of SNe Ic, and may come from different physical progenitors than SNe Ic. Indeed, other lines of evidence point in the same direction: SNe Ic-bl and GRBs prefer over-dense host galaxy environments , compared to other Stripped SNe, including SNe Ic \citep{kelly14}, and SNe Ic-bl (with and without GRBs) have lower environmental metallicities than those of SNe Ic \citep{modjaz08_Z,modjaz12_proc,sanders12,kelly12}.

 \subsection{Are SNe Ic-bl connected with LLGRBs different from those connected with HLGRB?} \label{llgrbvshlgrb}

Starting with the first case of the SN-GRB connection, namely \snbw, there have been so far four SN-LLGRBs observed, which as a class, may be 100$-$1000 times more common than HLGRBs (e.g., \citealt{cobb06,soderberg06_06aj,guetta07}). In addition to their low luminosities ($10^{46}- 10^{48}$ \ergs ), LLGRBs appear to be more isotropic and have relatively soft high-energy spectra compared to HLGRBs (e.g., \citealt{campana06,hjorth13,nakar15} and references therein). There have been suggestions that the high-energy emission of the GRB in these two groups\footnote{We note that there are also some GRBs whose $\gamma$-ray luminosity appear to be intermediate (e.g., SN 2012bz/GRB 120422A, Schulze et al. 2014).} may come from different mechanisms (e.g., \citealt{waxman07,bromberg11,nakar12,nakar15}), namely that LLGRB are from shock breakout of the collapsar jet from a low-density envelope or wind, and thus are more spherical because of that. 

In figure~\ref{vabs_GRB_fig}, we plot the measured SN absorption velocity as a function of isotropic gamma-ray luminosity ($left$) and of gamma-ray energy ($right$), showing both isotropic and beaming-corrected energies (taken from the compilation in \citealt{mazzali14}, see references therein, and also see \citealt{ryan15}). The velocities are based on the spectrum within $t_{Vmax}=\pm$2 days. As can be seen, when corrected for beaming angle, the intrinsic energies are similar between LLGRBs and HLGRBs and the only difference appears to be collimation, with LLGRBs being almost spherical. There is no clear correlation between the SN absorption velocities and the GRB isotropic energies and luminosities (for both cases, the Spearman coefficient is $\rho$= 0.2 with a low significance of 0.6), nor beaming-corrected $E_{\gamma}$  (the Spearman coefficient is $\rho$= 0.4 with a low significance of 0.3). A similar analysis correlating the SN peak optical luminosity with its GRB luminosity did not find a trend either \citep{hjorth13}. 


\section{Conclusions \& Future Directions}\label{conclusion_sec}

We have presented the first systematic investigation of spectral properties of explosions from massive stripped stars, which are sometimes accompanied by long-duration GRBs, in order to probe their explosion conditions and their progenitors. To that aim, we analyzed a total of \nspecanal\ spectra of \nSNGRB\ SNe Ic-bl with GRBs, \nSNIcbl\ SNe Ic-bl without observed GRBs, and \nSNIc\ SNe Ic. The data consist of published spectra of individual SNe from the literature as well as the densely time-sampled and homogeneous spectral data set of stripped SNe from \citet{modjaz14}. 

In order to quantify the diversity of the spectra of a specific sub-type, we constructed average spectra of normal SNe Ic, SNe Ic-bl with GRBs and SNe Ic-bl without GRBs, along with standard deviation and maximum deviation contours. 
We furthermore explore the ejecta velocities present in different kinds of SNe Ic using two different methods: 1) measuring \FeFive\ line velocities from their Doppler-shifts and 2) exploring the widths of the lines. We use a lot of care to derive robust error bars for all our measurements by constructing spectral uncertainty arrays for all our spectra from the SN spectra themselves and by propagating them into our final measurements, either via Monte Carlo resampling or by formal fitting using MCMC. In addition, we have developed a novel method for analyzing the blended features of SNe Ic-bl, where in particular the three lines of Fe~II (Fe~II~$\lambda$4924, Fe~II~$\lambda$5018, and \FeFive ) are blended into one broad feature in SNe Ic-bl. This novel method uses our prior constructed mean SN Ic spectra as templates for the appropriate phase and finds a best fit to the SN Ic-bl spectra, using MCMC to explore the full parameter space and to derive error bars.

The main results that we have obtained can be summarized as follows:

$\bullet$ SN~Ic 1994I is not a typical SN Ic, in contrast to common usage, while the spectra of \snbw\ are representative of mean spectra of SNe Ic-bl. While the spectra of \snbw\ lie close to the mean spectra of SNe Ic-bl with GRBs and well within one standard deviation, those of SN~1994I are close to the maximum deviation of the mean spectra of SNe Ic (especially at $t_{Vmax}$=10 days) at many wavelengths.

$\bullet$ For SN~Ic~1994I in particular, and for all SNe in general, the velocity obtained depends on which method was used to measure photospheric velocity, as different lines (\FeFive\ vs. a line commonly identified as \SiSix ) and different spectral synthesis codes (SYNOW vs. Mazzali's NLTE code) yield systematically different values. Furthermore, it is crucial to distinguish between two kinds of velocities (absorption velocity vs. line-width velocity) since they are not always correlated for the same SNe (e.g., PTF~12gzk and PTF~10vgv), and we encourage authors to be clear about which kind of velocity they are reporting, as well as to use the same method when comparing different SNe. We encourage future modeling of the spectra of SNe Ic-bl with a detailed parameter study to reproduce the observed parameter space of SNe Ic and Ic-bl, including asphericity in the ejecta and viewing angle effects.

$\bullet$ SNe Ic-bl (with and without GRBs) have \FeFive\ widths that are $\sim$9,000 \kms\ (median value) broader than those of SNe Ic. They also have  average \FeFive\ absorption velocities that are larger than those of SNe Ic by $\sim$9,000 \kms\ when excluding \snbh\, and by $\sim$13,000 \kms\ when including \snbh . The latter values are for the epoch of $V$-band maximum. 

$\bullet$ Somewhat surprisingly, we find that SNe Ic-bl with GRBs, on average, have higher ejecta velocities (by $\sim$6,000 \kms, if excluding \snbh ), as well as wider line widths (by $\sim$ 2,000 \kms) than SNe Ic-bl without observed GRBs, as traced by \FeFive. That is, the SN Ic-bl ``knows" whether it harbored a GRB beamed towards us. We can interpret these observations in two ways: the first is that those SNe Ic-bl without observed GRBs had lower-energy or shorter-duration jets that were stifled and could not break out of the star, but nevertheless imparted some of their energy to the stellar envelope and therefore accelerated it less than SNe with high-energy, visible jets. The other possibility is that SNe Ic-bl without observed GRBs did harbor GRBs that are highly beamed, but they were all beamed away from us, i.e., they were all off-axis GRBs. We disfavor the latter possibility since there is strong evidence that a number of the SNe Ic-bl in our sample did not harbor off-axis GRBs, and since many SN-GRBs are not highly beamed in general. In order to verify our hypothesis we suggest that the geometry of the ejecta of SNe Ic-bl without GRBs should be probed (e.g., via polarization studies or late-time spectra) to detect any other signs of a stifled jet inside SNe Ic-bl without observed GRBs.

$\bullet$ We search for differences in SN Ic-bl spectra for those connected with Low-Luminosity GRBs vs. those with High-Luminosity GRBs, as some theories suggest different physical mechanisms. While the sample sizes are still very small (with each group having $\sim$4-5 objects), we find no clear correlations between SN spectral properties and GRB energies, nor GRB luminosities.

$\bullet$ The absence of clear He lines in optical spectra of SNe Ic-bl cannot be due to the smearing out of the Helium lines by the high velocities present. While the lack of $^{56}$Ni mixing may in principle be responsible for "hiding" He lines, we show in our companion paper \citet{liu15} that the lack of $^{56}$Ni mixing is most likely not the only difference between SNe Ic and SNe Ib (also see \citealt{hachinger12,cano14_99dn}). These findings imply that the progenitor stars of SNe Ic-bl (with and without GRBs) are probably He-free, in addition to being H-free, which puts strong constraints on the stellar evolutionary paths needed to produce H-free and He-free progenitor stars at low metallicity. In particular, our He constraint supports short-period binary models for SNe Ic-bl, in line with other recent works (e.g., \citealt{kelly14}). In light of this, we encourage more studies of star formation conditions and of the shape of the initial mass function in galaxies similar to those host galaxies of SNe Ic-bl and GRBs, namely over-dense, low-mass, and low-metallicity galaxies, as well as theoretical investigations of removing both H and He in the potential binary systems of GRB progenitors.

$\bullet$ Overall, we encourage more observations of SNe Ic-bl without observed GRBs  - while we have similar numbers of SNe Ic-bl with and without GRBs in this work, the number of spectra for SN-GRBs is a factor of $\sim$3 larger than those of SNe Ic-bl without observed GRBs. This is also in contrast to the fact that many more SN Ic-bl are known to be without observed GRBs than with observed GRBs. In addition, deeper radio and X-ray limits are needed for SNe Ic-bl to test for the presence of low-luminosity GRBs and low-energy outflows. Recently, there have been some suggestions that SN-GRBs may have standardizable light curves \citep{cano14,li14}, which may have implications on any relationships between $^{56}$Ni production (which sets the luminosity) in the different GRB engine scenarios  and the physics of the SN ejecta, including its geometry and the observer's viewing angle.



All data products and codes are open access and open source and are available on our SNYU github websites\footnote{ \url{https://github.com/nyusngroup/SESNspectraLib} and \url{https://github.com/nyusngroup/SESNtemples}, also linked from our SNYU website, \url{http://www.cosmo.nyu.edu/SNYU/.} }, including the code and MCMC wrapper for measuring velocities in SNe Ic-bl, and the additional SNID templates for literature Stripped SNe, using the same methodology as in \citet{liu14}. The data of \citet{modjaz14} are available on the CfA SN archive and WISEREP.\footnote{www.weizmann.ac.il/astrophysics/wiserep}. \\


\acknowledgments
We would like to thank Stephane Blondin, Tom Matheson, Robert Kirshner, Saurabh Jha, Iair Arcavi, Dan Milisavljevic, Pat Kelly, Andrew MacFadyen, Tony Piro,  Bromberg, Sung-Chul Yoon, John Eldridge, Avishay Gal-Yam for helpful discussions. We thank S. Blondin for making available to us the codes from Blondin et al. (2012) to measure absorption velocities in normal SNe Ic and to produce mean spectra.

M. Modjaz is supported in parts by the NSF CAREER award AST-1352405 and by the NSF award AST-1413260. Y. Q. Liu and O. Graur are supported in part by the NSF award AST-1413260. Y. Q. Liu is in part supported by a \emph{NYU/CCPP James Arthur Graduate Award} and F. B. Bianco in part by the \emph{NYU/CCPP James Arthur Postdoctoral Fellowship}. 

Furthermore, this research has made use of NASA's Astrophysics Data
System Bibliographic Services (ADS), the HyperLEDA database and the	
NASA/IPAC Extragalactic Database (NED) which is operated by the Jet
Propulsion Laboratory, California Institute of Technology, under
contract with the National Aeronautics and Space Administration.  
This paper has made extensive use of the SUSPECT database (www.nhn.ou.edu/?suspect/)
and the Weizmann interactive supernova data repository
(www.weizmann.ac.il/astrophysics/wiserep)


\bibliographystyle{apj}

\clearpage



\appendix

\section{A. Measuring line velocities via the template-fitting method in SNe Ic-bl}\label{speclineID_sec}

\begin{figure*}[!ht]    
\vspace{-.8in}
\hspace{-.1in}
\includegraphics[width=0.45\columnwidth,angle=0]{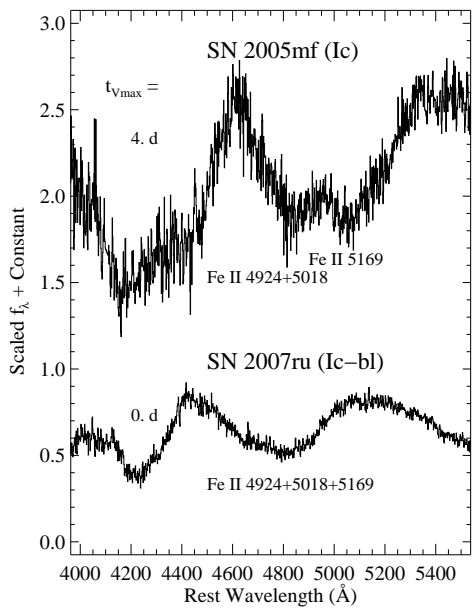}
\singlespace \caption{SN spectra of a SN Ic and a SN Ic-bl showing the wavelength region around the Fe II lines. }
\label{Fe-ID_fig}
\end{figure*}

\begin{figure*}[tb!]
\begin{center}
\centerline{
\includegraphics[width=.47\columnwidth,angle=0]{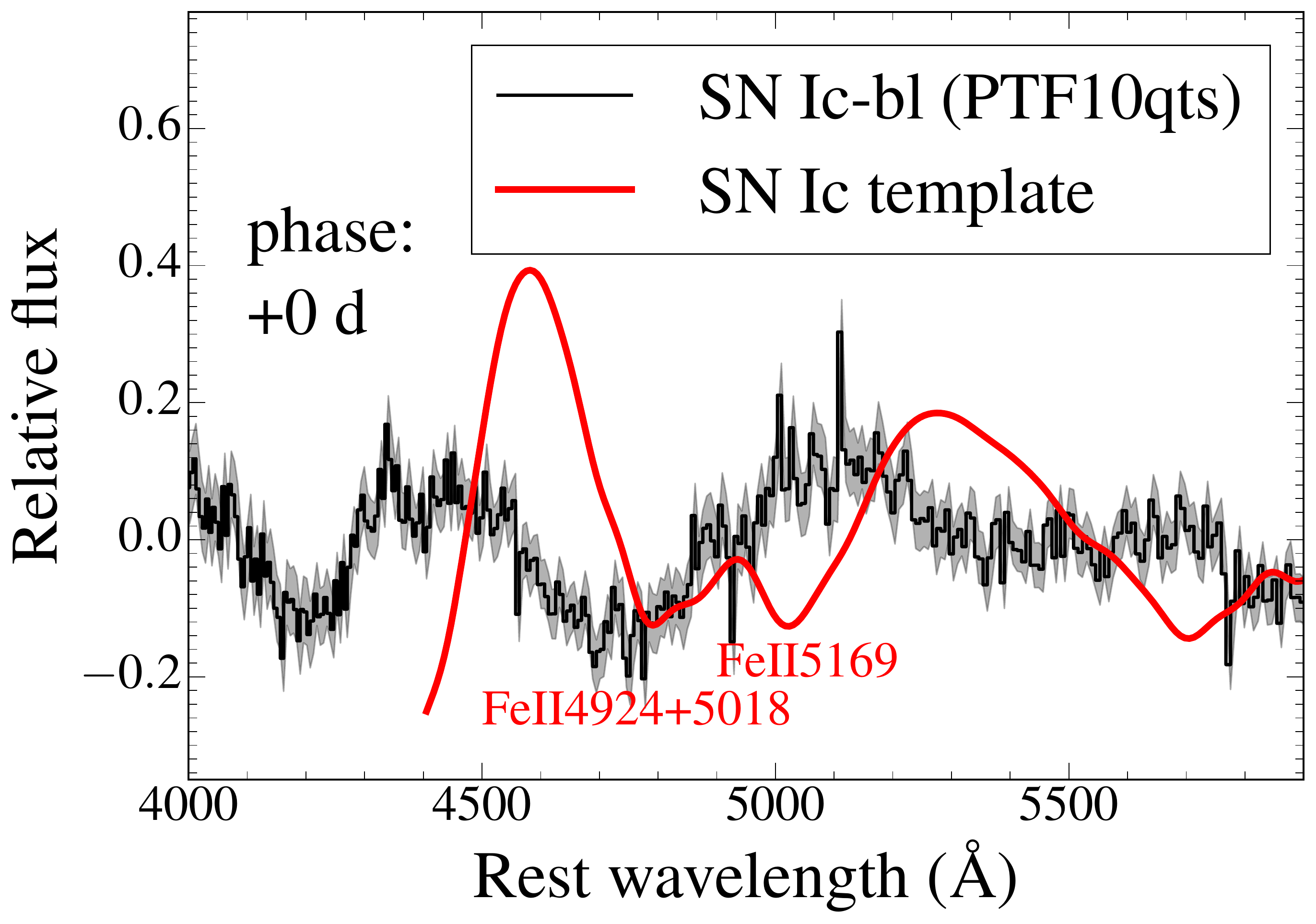} 
\hspace{.3in}
\includegraphics[width=.47\columnwidth,angle=0]{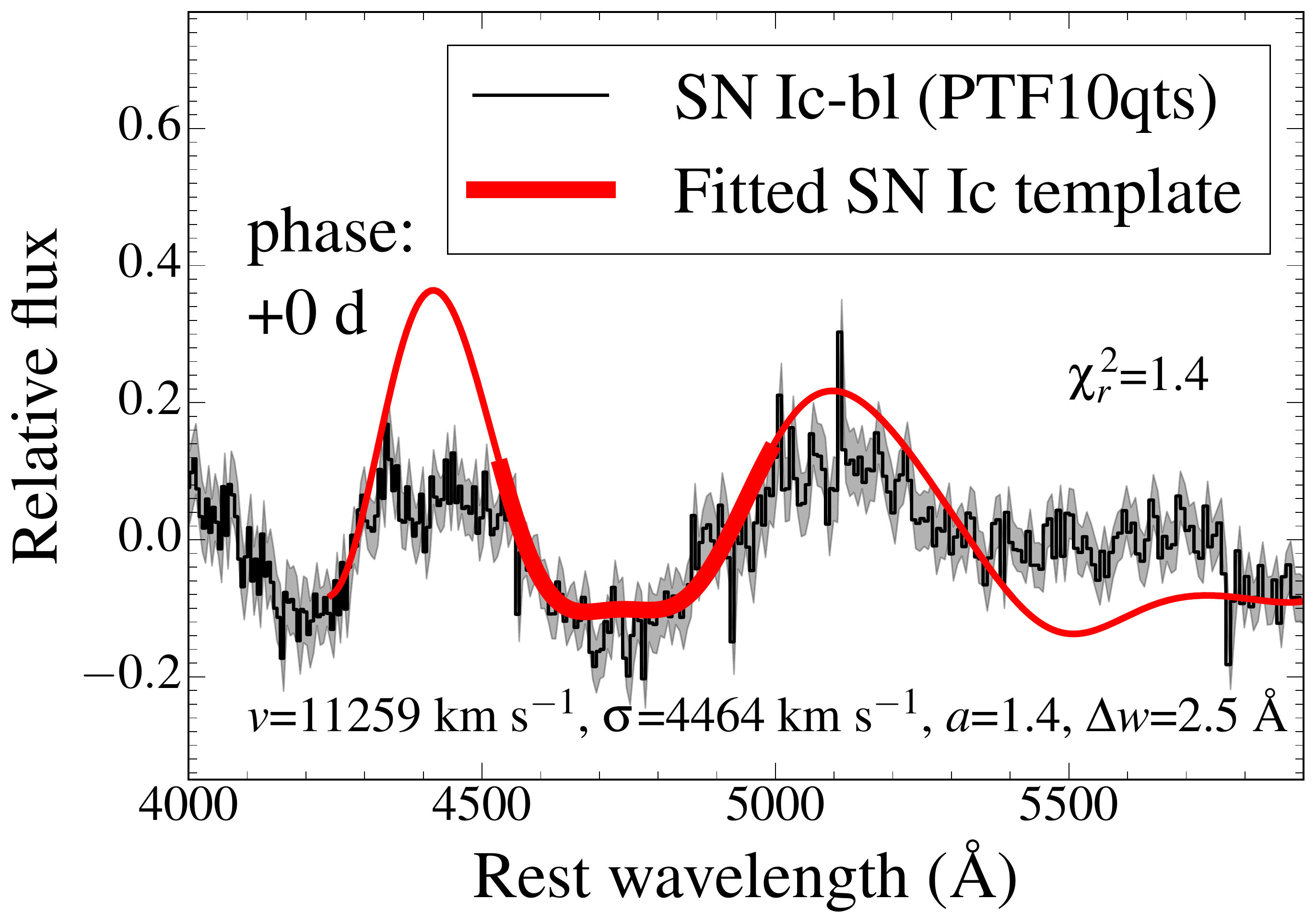} %
}
\caption{Example for convolution fitting of the Fe II feature for the spectrum of a SN Ic-bl (here: PTF10qts) at t=0 days. In both plots, the SN Ic-bl spectrum is in black with its error spectrum overlaid in grey. \emph{Left:} The SN Ic template spectrum (before the fit) is overplotted in red, and shows a ``W" feature between 4600 \AA\ and 5300 \AA\ due to the FeII triplet: the red absorption trough is due to Fe II $\lambda$5169, while the blue one is a blend of the other two Fe II lines. In the spectrum of the SN Ic-bl PTF10qts, the three Fe II lines are blended into one feature. \emph{Right}: The fitted SN Ic template spectrum is overplotted in red, with the 50th percentile confidence values and their error bars (corresponding to 50$^{th}-16^{th}$ and 84$^{th}-50^{th}$ percentiles respectively) determined by Markov-Chain Monte Carlo (MCMC) methods. The SN Ic template has been dopplershifted by the 50th percentile value of velocity parameter $v$ ($=11,260^{+270}_{-280} $ km s$^{-1}$), broadened by the Gaussian kernel with the 50th percentile value of sigma parameter $\sigma$ ($=4460^{+370}_{-320}$ km s$^{-1}$), and stretched along the $y$-axis with the best value of amplitude parameter $a$ ($=1.4^{+0.1}_{-0.1} $). The thick red line denotes the spectrum that is used for the fit with the best value of the wavelength-range parameter $\Delta w$ ($=2.5^{+4.2}_{-2.0} $ \AA), by which the initial wavelength range (between the two maxima) gets subtracted. Note that the final reported value (and its uncertainty) of the Doppler velocity parameter $v$ includes the contribution of the SN Ic template itself for each phase at which the fit is performed. Also shown on the plot is the reduced $\chi^2$ value for this fit. 
}
\label{fig_Icbl_Ic_fit}
\end{center}
\end{figure*}

The blueshift and width of specific spectral absorption features, such as Fe II $\lambda$ 5169, are used to track the photospheric velocity of SNe. For SN Ic spectra, it is straight-forward to identify absorption due to Fe II $\lambda$5169, while Fe II $\lambda\lambda$ 4924, 5018 lines are usually blended into a single feature (see Fig.~\ref{Fe-ID_fig}). For SNe Ic-bl, all three Fe II lines are heavily blended due to the high expansion velocities in the ejecta (see  Fig.~\ref{Fe-ID_fig} and~\ref{fig_Icbl_Ic_fit}). Thus, it is difficult to accurately identify Fe II $\lambda$ 5169 and measure its absorption velocity. There have been several attempts in the literature to solve this problem. \citet{lyman14}, for example, used Si II $\lambda$6355 instead, whenever they could not identify Fe II $\lambda$5169. However,  we find that the Fe II $\lambda$5169 absorption velocity is systematically higher than the Si II $\lambda$6355 measured using the same method (see Section~\ref{Fevabs_subsec} and Bianco et al, in prep). Thus, the two velocities cannot be used interchangeably to track photospheric velocity in order to calculate physical parameters for SNe. What makes it worse is that Si II $\lambda$6355 is sometimes blended with the Na I D lines into one absorption feature (e.g., PTF 10qts). Thus, it is at times difficult to identify Si II $\lambda$6355 as well. \citet{schulze14} analyzed spectra of SN -GRBs (many of the same spectra as in our SN-GRBs sample) and attributed the blended and broad feature due to the three Fe II lines to only Fe II $\lambda$5169. Thus, they derive systematically high velocities, up to $-$50,000 km s$^{-1}$.  \citet{pignata11} used SYNOW, a spectral synthesis code, to obtain velocities of Fe II $\lambda\lambda\lambda$4924, 5018, 5169 for SN 2009bb. Different spectral synthesis codes, however, yield different velocities for the same SN using different methods, as shown in Section~\ref{specline_94Isubsec}.

We developed a data-driven and consistent method to identify Fe II $\lambda$5169 and measure its velocity in the blended spectra of SNe Ic-bl:  we used a template fitting approach in which we found the best fit between a convolved and blueshifted SN Ic template and the SN Ic-bl spectrum under consideration, at similar phases, for the wavelength region of the feature under consideration. This is similar to what we did in Section~\ref{he_subsec} for the whole spectrum, but now it is a formal fit, which we implement via Monte Carlo Markov Chain (MCMC) methods. The fits that this procedure yields are generally good, as indicated by the $\sim$ unity values for the reduced $\chi^2$  values.
 
Our SN Ic template spectra are the mean SN Ic spectra of Section~\ref{averagespec_sec}, but with a different time spacing: now we construct them for every two-day-phase interval (as opposed to five-day-intervals for the mean spectra of Section \ref{average_subsec}), in the range between $t = -10$ and $t= +72$ days with respect to maximum light (i.e, $t = -10, -8, -6, -4$, etc. days). Each mean spectrum includes SN Ic spectra in 5-day bins centered around the middle, i.e. "target" phase. For example: in order to measure the photospheric velocity for a SN Ic-bl spectrum at phase $t=$0 days, we use the mean SN Ic spectrum constructed using SN Ic spectra between phases $-2.5$ and 2.5 days and centered at phase $t=0$ days. Thus, we note that template spectra at adjacent phases may not be independent from each other, but this aspect should not influence our results. For measuring the velocity of a SN Ic-bl spectrum at phase $t=-2$ days, we use the mean SN Ic spectrum centered at $t=-2$ days (constructed from SN Ic spectra between $-$4 and 0 days) for template fitting.
  Each mean spectrum includes no more than one spectrum per SN Ic, in order to avoid biasing the mean spectrum towards better-observed SNe. If two or more spectra of the same SN Ic are available within the phase range (within $\pm2.5$ days of the target phase), we choose the one nearest to the target phase. If there is still more than one spectrum that satisfies the above condition, we use the one that covers the largest optical wavelength range, or the one with the highest signal-to-noise ratio. Each mean spectrum contains spectra of $N$ SNe, where $N\ge3$, except the ones at phases of $t=+$62 and $+$64 days. Thus, for SNe Ic-bl spectra with phases between $t=+$61 and $+$65 days, we use the mean spectrum with the nearest phase --- either the one at phase $t=+60$ days or the one at phase $t=+66$ days --- to replace mean spectra at phases 62 and 64. Since SN Ic-bl spectra have broader features and higher velocities than those of SNe Ic, we convolve the template with a Gaussian kernel (see Section~\ref{he_subsec}) and doppler blueshift it using the relativistic Doppler formula.

There are four parameters in our fit: a parameter for the width of the convolution Gaussian ($\sigma$), a velocity parameter ($v$) for the doppler shift, an amplitude parameter ($a$), and a wavelength-range parameter ($\Delta w$), by which we add to or subtract, times two, from the initial guess for the wavelength range of the SNe Ic-bl spectrum over which we will fit the template. The initial guess of the wavelength range is based on finding the local maxima on the blue and the red side of the feature and using the range in-between. We choose to use a single parameter instead of two (i.e., one for each end) to control the range because we want to limit the dimensionality of our parameter space, as a higher dimensional parameter space induces higher uncertainty in the parameter distributions, and requires more computational power. We use the following priors on our parameters: uniform prior for absorption velocity with respect to the SN Ic template (0$-$20,000\kms, where positive values refer to blueshifts, i.e., negative velocities ), $\sigma$ (0$-$20,000 \kms, where positive values refer to wider kernels), and amplitude (0$-$3).  We used a gaussian prior for the wavelength-range parameter $\Delta w$, where we centered the Gaussian on zero and used 100 \AA\ for the 99.7th percentile region. We note that these choice are only valid to up to $t_{Vmax}<$ 60 days, as afterwards the SNe Ic-bl spectra become too similar to the SN Ic template spectra.



\begin{figure*}[tb!]
\begin{center}
\centerline{
\includegraphics[width=0.62\columnwidth,angle=0]{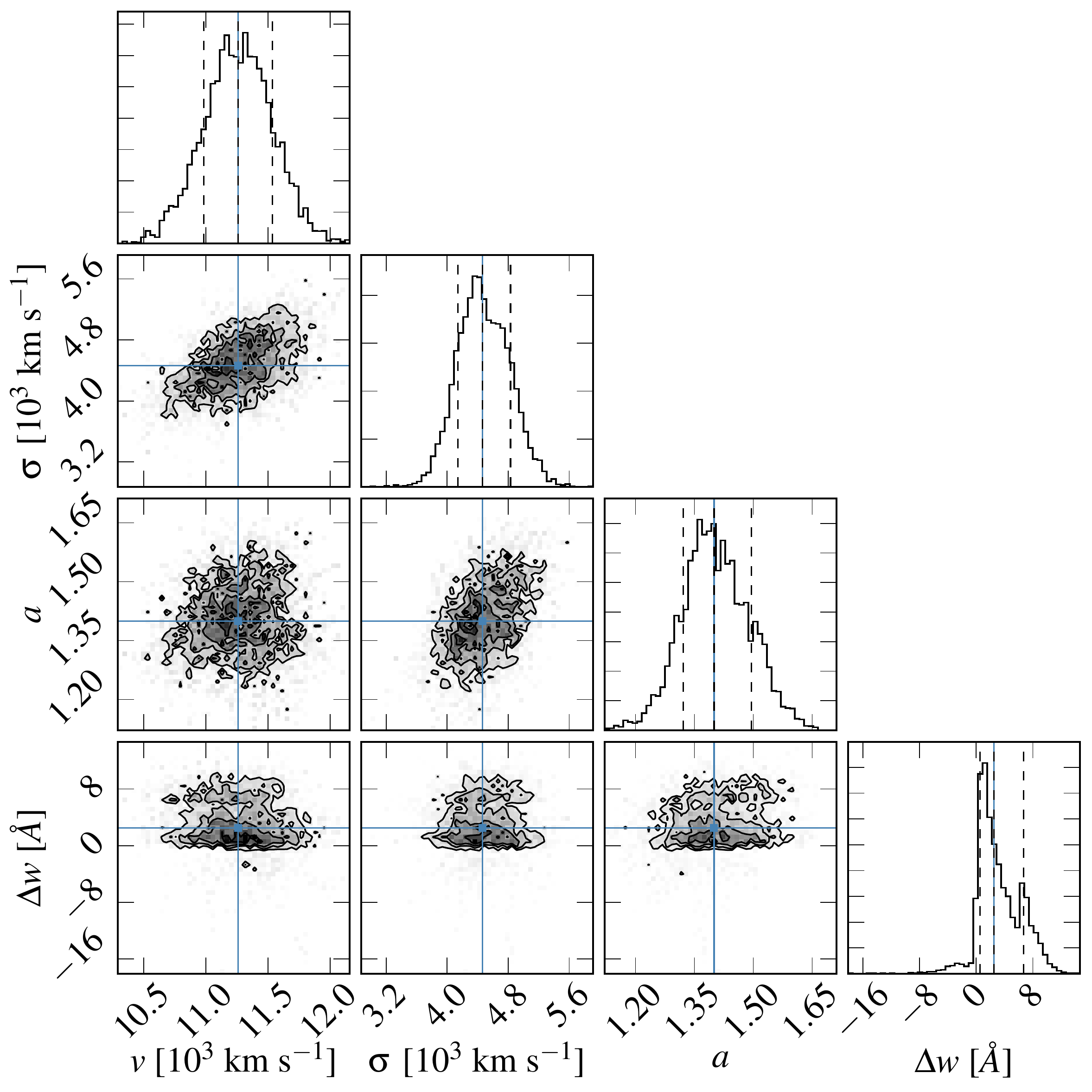} 
}
\caption{One- and two-dimensional projections of the posterior MCMC probability distributions for the spectral fit parameters for PTF~10qts at $t_{Vmax}$=0, shown in Fig.~\ref{fig_Icbl_Ic_fit}. This figure was made with \textbf{corner.py} \citep{foreman-mackey16}.}
\vspace{-.1in}
\label{fig_mcmc}
\end{center}
\end{figure*}

In summary, our likelihood function is:
\begin{multline}
\mathrm{ln}~p(f_{Icbl}|f_{Ic},err_{fIcbl}, \sigma,v,a,\Delta w)= -0.5\times \sum_{x=wbi+\Delta w}^{wri - \Delta w} \left \{  \frac{\left \{ f_{Icbl}(x)-[a\times D(v)\times (G(s)\ast f_{Ic})](x)\right \}^2}{err_{fIcbl}(x) ^2\times (N-4)} + \ln [err_{fIcbl}(x)^2]\right \},
\label{equ_like}
\end{multline}
where $f_{Icbl}(x)$ is the flux of a SN Ic-bl spectrum at wavelength $x$, $D(v)$ is the relativistic doppler shift at velocity $v$, $G(s)$ is the Gaussian function with width $\sigma$, $f_{Ic}$ is the flux of the SN Ic mean spectrum, $err_{fIcbl}$ is the uncertainty of $f_{Icbl}$, $N$ is the number of data points between wavelength $wbi+\Delta w$ and $wri-\Delta w$, and 4 represents the number of parameters in our model. We sample the posterior distribution function via the \textbf{emcee} package \citep{foreman-mackey13} which implements an affine-invariant ensemble Markov chain Monte Carlo (MCMC) sampler (Goodman \& Weare 2010). The velocity parameter $v$ and its error bar are based on its marginalized distribution: the Doppler velocity parameter $v$ corresponds to the 50th percentile in the marginalized $v$ distribution (i.e., median), while its error bars correspond to the 50$^{th}-16^{th}$ and 84$^{th}-50^{th}$ percentiles of the marginalized distribution of $v$ (Fig.~\ref{fig_mcmc}). We compute the final Fe II $\lambda$5169 velocity for SNe Ic-bl by adding the Fe II $\lambda$5169 absorption velocity of the SN Ic template to the velocity parameter value from the fit. We compute the final uncertainty in the Doppler velocity parameter $v$ by adding in quadrature the MCMC-derived uncertainty to that based on the velocity error in the corresponding SN Ic template. The velocity error in the SN Ic template is based on the fact the template spectrum is generated by spectra of different SNe Ic and at slightly different phases - thus the template velocity error was determined by taking the standard deviation of velocities measured in the individual SN spectra that were used to generate the SN template spectrum. Note that a potential mismatch between the phase of the SN Ic-bl spectrum and that of the mean SN Ic spectrum that was used for the fit does not impact the final absorption velocity measured for the SN Ic-bl spectrum.

In Figure~\ref{fig_Icbl_Ic_fit}, we demonstrate our template-fitting method for the SN Ic-bl PTF10qts at phase $t_{Vmax} = 0$ days, using the SN Ic mean spectrum at phase $t_{Vmax} = 0$ days as a template. Since the SNe Ic mean spectra are constructed using flattened spectra, we use the flattened version of the SNe Ic-bl spectra as well. We note that for SN~1998bw, its blue Fe trough is stronger than the one in the SN Ic template, at all phases. The same applies to SN~2002ap, but only for up to $t_{Vmax} = 10$ days. 

We tested our template-fitting method by applying it to SNe Ic spectra and comparing the resultant Fe II $5169$ velocities with those from line identifications. We find that the Fe II $5169$ velocities from the two methods are consistent within their error bars. Thus, we can be confident that using the template-fitting method, which we use for SNe Ic-bl does not introduce a systematic bias in the velocity measurements. 
We note that for SN Ic 2013dk, the three iron lines were blended into one feature, such that we used the template-fitting method to measure its absorption velocity.  Our code to measure absorption and width velocities for SNe Ic-bl (and for any other SNe with high line-blending) is available on github under \url{https://github.com/nyusngroup/SESNspectraLib} .

\section{B. Absorption velocities of individual SNe}\label{vabs_sec}
\begin{figure}[tb!]
\begin{tabular}{c}
\vspace{-.5in}
\includegraphics[width=.6\columnwidth,angle=0]{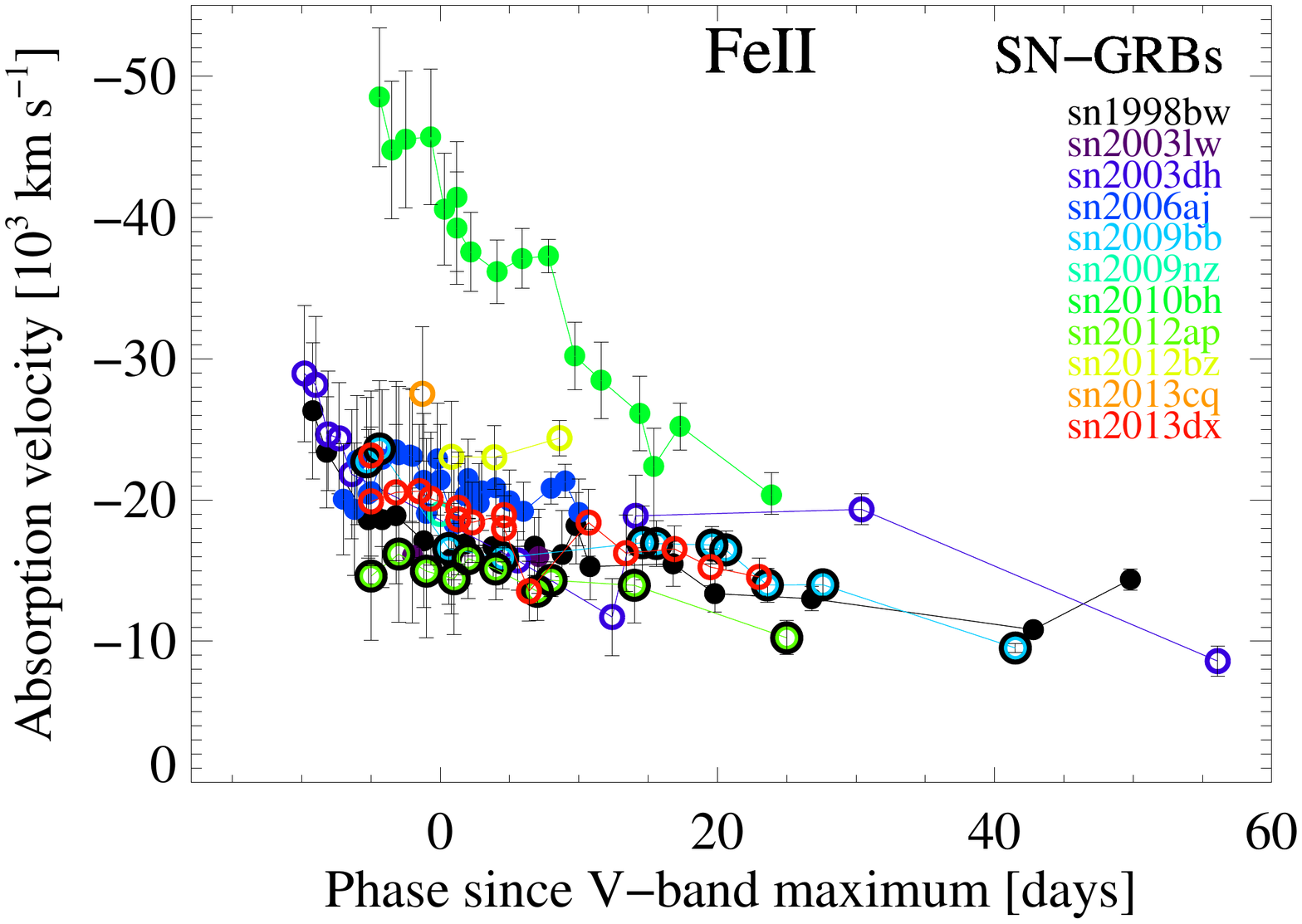} \\
\vspace{-1.in}
\includegraphics[width=.6\columnwidth,angle=0]{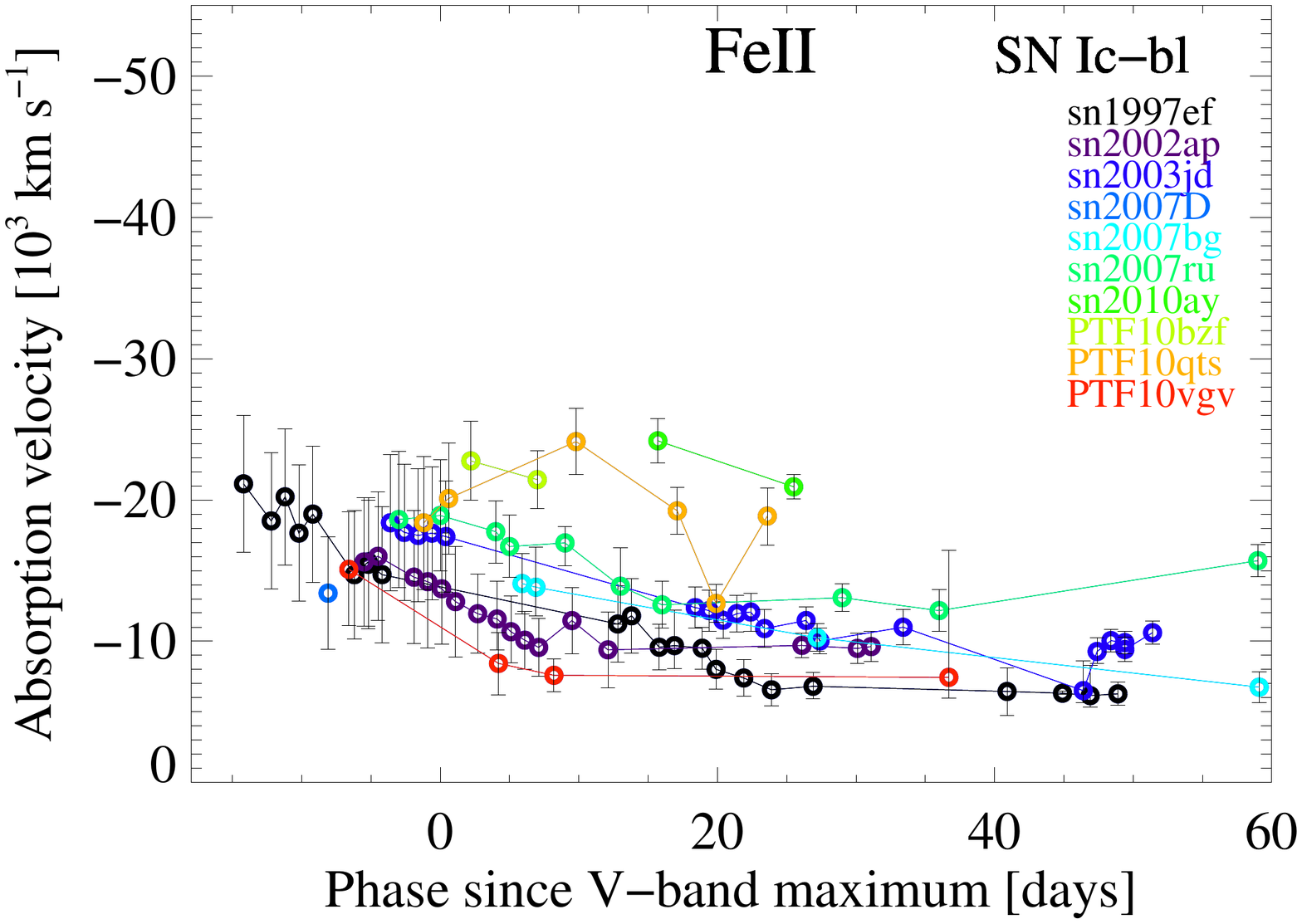} \\
\includegraphics[width=.6\columnwidth,angle=0]{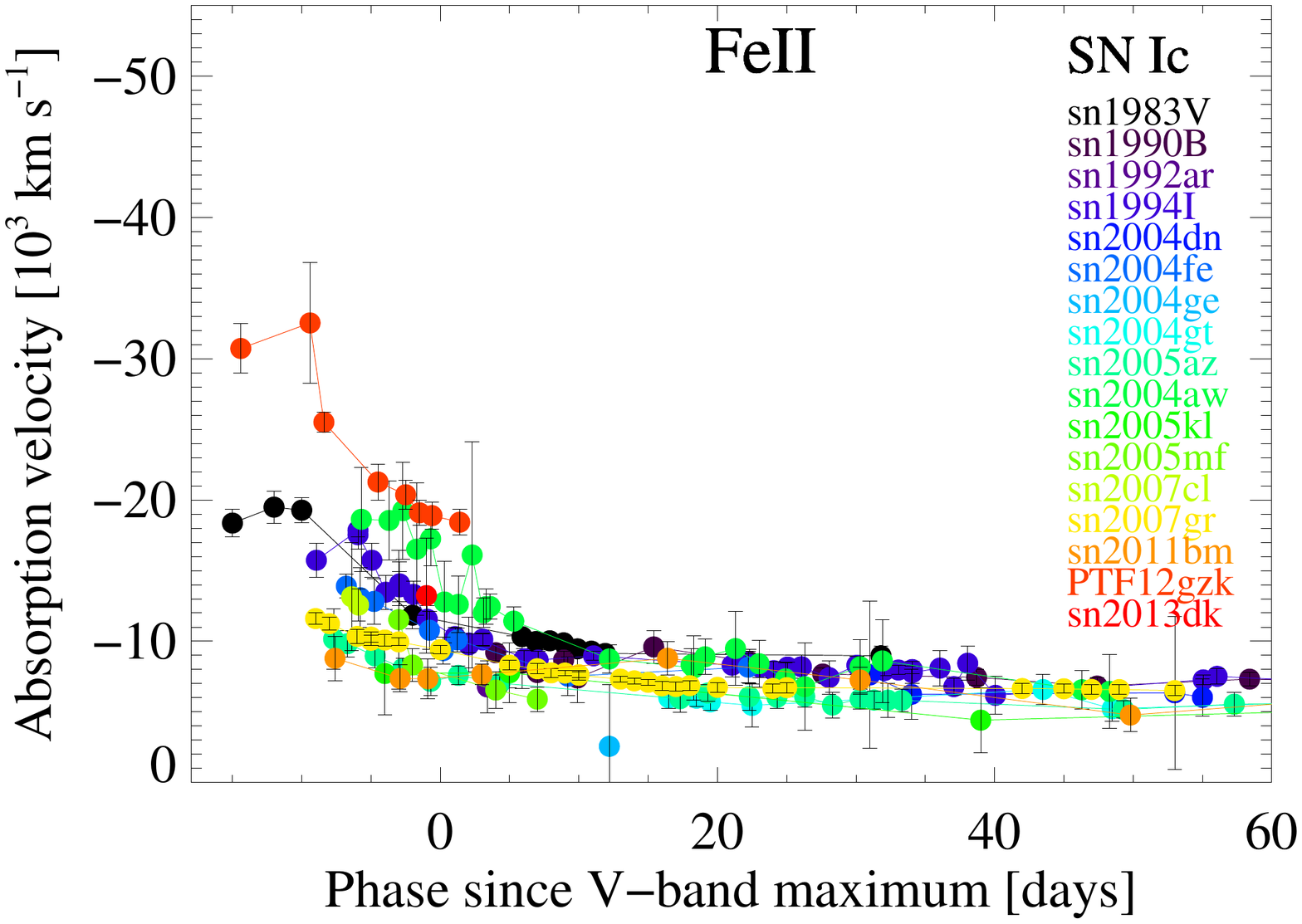} \\

\end{tabular}
\caption{\FeFive\ absorption velocities for individual SNe, grouped according to SN type. For the top plot that shows SN-GRBs, we also distinguish between SN Ic-bl connected with LLGRBs (filled circles, SNe 1998bw, 2003lw, 2006aj, 2010bh) vs. opposed to the rest of the SNe Ic-bl (open circles). The latter include those that are connected with HLGRBs (SNe 2003dh, 2009nz, 2013cq) and with intermediate-luminosity GRBs (SNe 2012bz, 2013dx) and those that did not have an accompanied GRB, but exhibited very strong radio-emission, suggesting a relativistic engine (SNe 2009bb, 2012ap - indicated by an extra black circle around them). }

\label{fig_absvels_sntypes}
\end{figure}

While we presented the bulk properties of SNe in the above sections, readers may be interested in the velocity evolution of individual SNe. Thus, in FIg.~\ref{fig_absvels_sntypes}, we present the \FeFive\ absorption velocities of individual SNe in three panels for the different SN subtypes. For the plot that shows SN-GRBs, we also distinguish between SN Ic-bl connected with LLGRBs vs. SNe Ic-bl connected with HLGRBs, as well as the radio-relativistic SNe 2009bb and 2012ap.

\end{document}